\documentclass[aps,prd,superscriptaddress,twoside,twocolumn,nofootinbib,10pt,showpacs,floatfix]{revtex4-1}

\usepackage{amsmath,amssymb}
\usepackage{graphicx,bm}
\usepackage{ulem}
\usepackage{slashed}
\usepackage{dcolumn}
\usepackage{epsfig}
\usepackage{multirow}
\usepackage[arrowdel]{physics}
\newcommand{\pvec}[1]{\vec{#1}\mkern2mu\vphantom{#1}}

\begin{document}

\preprint{}

\title{Origin of energy shift in kaonic atom and kaon-nucleus interaction}

\author{Yutaro Iizawa}

\email{iizawa-yutaro@ed.tmu.ac.jp}
\affiliation{Department of Physics, Tokyo Metropolitan University, Hachioji 192-0397, Japan}
\affiliation{Department of Physics, Tokyo Institute of Technology, 2-12-1 Ookayama, Megro, Tokyo 152-8551, Japan}

\author{Daisuke Jido}
\affiliation{Department of Physics, Tokyo Institute of Technology, 2-12-1 Ookayama, Megro, Tokyo 152-8551, Japan}
\affiliation{Department of Physics, Tokyo Metropolitan University, Hachioji 192-0397, Japan}

\author{Natsumi Ikeno}
\affiliation{Department of Life and Environmental Agricultural Sciences, Tottori University, Tottori 680-8551, Japan}

\affiliation{Departamento de F\'{\i}sica Te\'orica and IFIC, Centro Mixto Universidad 
de Valencia-CSIC Institutos de Investigaci\'on de Paterna, Aptdo. 22085, 
46071 Valencia, Spain}

\author{Junko Yamagata-Sekihara}
\affiliation{Department of Physics, Kyoto Sangyo University, 436 Motoyama Kamigamo, Kita-ku, Kyoto 603-8555, Japan}

\author{Satoru Hirenzaki}
\affiliation{Department of Physics, Nara Women's University, Nara 630-8506, Japan}

\date{\today}

\begin{abstract}
  The $K^-$-nucleus optical potential is revisited to investigate its global feature phenomenologically. It is a puzzle that the energy shift is found to be repulsive in all of the observed kaonic atom, although the $K^-N$ interaction is known to be so attractive as to form the $\Lambda(1405)$ resonance. To solve this puzzle, we examine the $K^-$ optical potential in the linear density approximation and determine the potential parameters of each kaonic atom so as to reproduce the observed energy shift and absorption width. We find two types of the potentials. One potential has a so large real part as to provide nuclear states with the same quantum number to the atomic state in the last orbit. The level repulsion between the atomic state and the nuclear states takes place due to their mixing, and it makes the atomic state shifted repulsively. The other type of the potential has a large imaginary part and the imaginary part works repulsively for atomic states. We find that only the latter solution reproduce a wide of the observed data, and thus is realized as a $K^-$-nucleus potential for kaonic atom. {In the linear nuclear density optical potential, the picture that the repulsive shifts in the atomic states stem from the existence of the nuclear states does not globally stand up.  This implies that the $K^-$-nucleus optical potential should have a large imaginary part. We examine some nonlinear density effects and find that the conclusion does not change. We also confirm that the conventionally known optical potentials are categorized into the latter type of the potential.}
\end{abstract}


\maketitle

\section{Introduction}
The study of the in-medium properties of kaon has attracted continuous attention and is one of the fundamental topics of contemporary nuclear physics. Kaon is the lightest meson with strangeness and one of the Nambu-Goldstone bosons of the spontaneous symmetry breaking of the SU(3) chiral symmetry. Investigating the in-medium properties of kaon, one expects to reveal the role of the strangeness in the nuclear medium. The kaon-nucleus interaction is one of the key quantities in the investigation of the in-medium kaon and can be extracted from kaonic atoms, which are bound states of a $K^-$ and a nucleus attracted mainly by the Coulomb interaction. The hadronic interactions shift the energy of the atomic state from that of the pure Coulombic bound state and provides the absorption widths. The energy shift and the absorption width are extracted experimentally through the X-ray spectroscopy. The $K^-$ meson that stops around a nucleus is captured in an atomic orbit, is deexcited by emitting X rays, and is absorbed into the nucleus. In experiments, the emitted X rays are observed.

So far, a lot of works have been done for the extraction of the optical potential of $K^-$ in nucleus from the experimental data of kaonic atoms \cite{Friedman:1994hx,Batty:1997zp,Friedman:2007zza,Batty:1981qi,Friedman:1999rh,Friedman:2012pc,Friedman:2012qy,Gal:2013vx,Barnea:2006kv,Cieply:2001yg}. For instance, in Ref.~\cite{Friedman:1994hx}, a global fit of the optical potential to the observed experimental data for the shift and width was performed, and the so-called phenomenological deep potential was obtained together with a shallower potential with a linear density approximation. Theoretical calculations have been also done in Refs.~\cite{Koch:1994mj,Schaffner:1996kv,Waas:1996xh,Waas:1996fy,Waas:1997pe,Ohnishi:1997pt,Lutz:1997wt,Tsushima:1997df,Sibirtsev:1998vz,Ramos:1999ku,Hirenzaki:2000da,Baca:2000ic,Tolos:2000fj,Tolos:2002ud,Korpa:2004ae,Tolos:2006ny,Yamagata:2006sv, Tolos:2008di,Weise:2008aj,Cieply:2011fy,Sekihara:2012wj,YamagataSekihara:2012rr}, for instance, using the chiral unitary approach \cite{Hirenzaki:2000da}. The theoretical calculations support the shallower optical potential. 
We note that the in-flight $( K^{-} , N )$ reaction was proposed in Ref.~\cite{Kishimoto:1999yj} as another way to extract the $K^{-}$ optical potential. The reaction experiment with carbon targets was performed as reported in Refs.~\cite{Kishimoto:2003jr, Kishimoto:2007zz}, and theoretical calculations for the in-flight reaction were also done in Refs.~\cite{Ikuta:2002kr,Yamagata:2005ic,Yamagata:2006sm,Yamagata:2007cp,YamagataSekihara:2008ji,Koike:2007iz,Magas:2009dk}.
A recent study in this line is found in Ref.~\cite{Ichikawa:2017lmh} carried out as a by-product of a pilot run of the J-PARC E05 experiment.

In this paper, apart of the big issue for the depth of the optical potential, we investigate a global feature of the optical potential by arising a question what is the origin of the repulsive shift seen in all of the observed kaonic atoms. It is well-known that the hadronic $\bar KN$ interaction is so attractive as to form $\Lambda(1405)$ as a bound state of $\bar KN$ with $I=0$  \cite{Kaiser:1995eg,Oset:1997it,Dalitz:1967fp,Hyodo:2011ur}. Nevertheless, in the kaonic atoms, the energy shift induced by the strong interaction is upwards repulsively. {In other words, the scattering length of the $\bar{K}N$ with $I=0$ is known to have repulsive nature. This is a consequence of the presence of the $\Lambda(1405)$ resonance below the $\bar KN$ scattering threshold, which induces the level repulsion due to the mixing of the $\Lambda(1405)$ resonance and the $\bar KN$ scattering states.} There are two possible solutions for this puzzle. One is the level repulsion between the atomic and nuclear states {as seen in the $\bar KN$ system.} When the optical potential is so strong as to provide a nuclear state closed to the atomic state, with coupling of these states level repulsion takes place and the atomic state is pushed away upwards. The other solution is an effect of a large imaginary part of the optical potential ${\rm Im}V_{\rm opt}$, which works as repulsive interactions. The imaginary part provides nuclear absorption in 1st order perturbation, while in the 2nd order perturbation it gives repulsive interaction. The aim of our study is to find which mechanism is realized in the actual kaonic atoms.

This paper is organized as follows. In Sec.~\ref{sec:2} we introduce two possible origins of the energy shift in kaonic atom. In Sec.~\ref{sec:3} we formulate the equation of motion of kaonic atom with the optical potential. In Sec.~\ref{sec:4} we determine the parameters of the linear nuclear density optical potential. We find two types of the situations and examine their universality. We examine their universality. In Sec.~\ref{sec:5} we investigate other potentials which are accepted as $K^-$ optical potentials. In Sec.~\ref{sec:6} we summarize the results of this paper.

\section{Two possible origins of the energy shift}
\label{sec:2}
In this paper, we focus on the origin of the energy shift in kaonic atom.
The energy shift is defined by the following formula:
\begin{align}
  \Delta E = { E_B -E_c}
\end{align}
where $E_c$ is the postulatory binding energy of the kaon and the atomic nucleus calculated by considering only the Coulomb potential and $E_B$ is the real part of the authentic binding energy including both the effects of the strong and electromagnetic interactions. 
 If $\Delta E > 0$ ($\Delta E < 0$) the strong interaction effect is attractive (repulsive).

The strong interaction effects on kaonic atoms can be seen in the energy shift from the pure Coulomb energy spectrum and the absorption width. The energy shift is found to be repulsive in all observed kaonic atoms. However the strong interaction between $K^-N$ is known to be attractive because $\Lambda(1405)$ is considered to be the $\bar{K}N$ bound state and therefore $K^-$-nucleus potential is considered to be attractive as well as $K^-N$ interaction. Hence we study the origin of the repulsive shift in kaonic atoms from the attractive strong interaction.

We have two candidates of the origin of the repulsive shift. One is an effect of strong attractive interactions. If the attraction is so strong that some nuclear bound states are generated under the atomic states, the atomic states are repelled  upwards by these bound states as a consequence of level repulsion. This mechanism is also seen in the $\bar{K}N$ system. The scattering length of the $\bar{K}N$ with $I=0$ is known to have repulsive nature even though the $\bar{K}N$ interaction is attractive. This puzzle is solved by the presence of the $\Lambda(1405)$ resonance below the threshold of $\bar{K}N$. The coupling of $\Lambda(1405)$ and $\bar{K}N$ scattering states causes the level repulsion and consequently the $\bar{K}N$ scattering length is seen as if $\bar{K}N$ were repulsively interacting. In this case the real part of the optical potential may be more significant for the repulsive shift than the imaginary part.

The other is an effect of a large imaginary part of the optical potential. It works repulsively, because, for example, the 2nd order perturbation of the large imaginary part

\begin{align}
  \label{shift}
  E_{n}^{(2)} = \sum _{m \neq n} \frac{\matrixel{ \phi_{n}}{i\Im V_{\mathrm{opt}}}{\phi_{m}}^2 }{E_{n}-E_{m}}
\end{align}
may give a positive value. Therefore if the optical potential has a large imaginary part, it gives the repulsive energy shift, which is irrelevant to level repulsion under attractive real part. In this case the imaginary part of the optical potential may be larger than the real part. In other words, the large imaginary part suppresses the atomic wavefunction inside the nucleus. Consequently the normalized wavefunction is pushed away to the outside of the nucleus. Therefore the large imaginary part plays an important role in the repulsive shift as discussed along the context of deeply bound antiproton atoms in Ref.\cite{Friedman:2007zza}.

We study the global feature of $K^-$-nucleus potential in the following approach : Firstly, we determine optical potential parameters for each observed kaonic atom by solving Klein-Gordon equation so as to reproduce the experimental data of the energy shift and the absorption width. Next, we confirm whether the obtained potential parameters describe also the other kaonic atoms by calculating the energy shifts and the absorption widths of various kaonic atoms using the parameters fitted by each kaonic atom and checking whether the calculated values are consistent with the experimental data.

\section{Model of kaonic atom}
\label{sec:3}
In this section, we explain our formulation to calculate the binding energy of a kaonic atom. We solve the Klein-Gordon equation for the relative motion of $K^-$ and a nucleus

We formulate the equation of motion with the Coulomb potential $V_c(r)$ and the optical potential $V_{\mathrm{opt}}(r)$. We study the optical potential between kaon and nucleus by solving the Klein-Gordon equation
\begin{align}
  \label{KG}
  \qty[-\qty{E-V_c(r)}^2-\laplacian + \mu^2 + 2 \mu V_{\mathrm{opt}}(r)]\phi(\vec{r})=0,
\end{align}
with the Coulomb potential $V_c(r)$ of the electromagnetic interaction and the optical potential $V_{\rm opt}(r)$ of the strong interaction. The complex energy $E$ may be written as $E=\mu - E_B - i \Gamma/2$ with the reduced mass $\mu$, the binding energy $E_B$ and the absorption width $\Gamma$. We assume spherical potentials for $V_c(r)$ and $V_{\rm opt}(r)$, and the wavefunction $\phi (\vec r)$ can be decomposed as $\phi(\vec r) = R_{n\ell}(r)Y_\ell^m(\Omega)$ with the radial wavefunction $R(r)$ and the spherical harmonics $Y_\ell^m(\Omega)$.

In order to see a global feature of the optical potential, we construct the $K^-$-nucleus optical potential simply under the linear nuclear density approximation. The $K^-$-nucleus optical potential $V_{\mathrm{opt}}(r)$ which is assumed to be proportional to nuclear density $\rho_N(r)$ is given by
\begin{align}
  \label{strong}
  V_{\mathrm{opt}}(r) = -(V_0+iW_0)\frac{\rho_N(r)}{\rho_0}
\end{align}
where $V_0$ and $W_0$ are parameters of the strength of the strong interaction. These parameters are determined by two experimental values, which are the energy shift and the absorption width for each kaonic atom. We assume that the strength of the optical potential does not depend on nucleus.
We use the Woods-Saxon form for the nuclear density $\rho_N(r)$ distribution :
\begin{align}
  \rho_N(r) & =\frac{\rho_0}{1+\exp[(r-R_B)/a]}
\end{align}
where $R_B$ and $a$ are the nuclear radius and diffuseness parameters. The parameters are taken from Refs.~\cite{DeJager:1974liz, DeJager:1987qc} and summarized in Table \ref{tab:data}.

The imaginary part of the optical potential produces nuclear absorption.
We assume that the real part of the optical potential works attractively because of the fact that the $K^-N$ interaction is attractive.

For the Coulomb potential $V_c(r)$, the charge distribution of the nucleus is taken into account as
\begin{align}
  \label{Coulomb}
  V_c(r)=-\frac{e^2}{4\pi} \int \frac{\rho_p(r')}{|\vec{r}-\pvec{r}'|} \dd[3]r',
\end{align}
where $\rho_p(r)$ is the proton density. We assume that the same density distribution as $\rho_N(r)$ for the proton density $\rho_p(r) = \rho_{p0} / (1+\exp[(r-R_B)/a])$ with $\rho_{p0}$ normalized as
\begin{align}
  \int \rho_p(r) \dd [3] r = Z,
\end{align}
with the proton number $Z$. The Coulomb energy $E_c$ appear in Eq.~(\ref{shift}) is obtained by $V_\mathrm{opt} = 0$.

In this paper, we do not consider the effect of vacuum polarization. We have confirmed that this effect to the energy shift and the absorption width is at most 5 percents and these value are smaller than the experimental errors.

\begin{table}
  \caption{Density parameters, the nuclear radius $R_B$ and the diffuseness $a$, used in this paper. The values are taken from Refs.~\cite{DeJager:1974liz, DeJager:1987qc}}
  \label{tab:data}
  \begin{ruledtabular}
    \begin{tabular*}{8.6cm}{@{\extracolsep{\fill}}ccc}
      Nucleus     & $R_B$[fm]   & $a$[fm]  \\ \hline
      Mg         & 2.980       & 0.551    \\
      Al         & 2.840       & 0.569    \\
      Si         & 2.930       & 0.569    \\
      P          & 3.078       & 0.569    \\
      S          & 3.165       & 0.569    \\
      Cl         & 3.395       & 0.569    \\

      Co         & 4.080       & 0.569    \\
      Ni         & 4.090       & 0.569    \\
      Cu         & 4.200       & 0.569    \\

      Ag         & 5.300       & 0.532    \\
      Cd         & 5.380       & 0.563    \\
      In         & 5.357       & 0.563    \\
      Sn         & 5.300       & 0.583    \\
    \end{tabular*}
  \end{ruledtabular}
\end{table}

\begin{table}
  \caption{Observed data of the energy shift $\Delta E$ and the absorption width $\Gamma$. The values are taken from Refs.~\cite{Backenstoss:1972yg,Batty:1979zr,Barnes:1974iu}}
  \label{tab:Katomdata}
  \begin{ruledtabular}
    \begin{tabular*}{8.6cm}{@{\extracolsep{\fill}}ccccc}
      Atom       &Transition        & $\Delta E$[keV]   & $\Gamma$[keV]  & Ref.\\ \hline
      Mg         &$4f$$\rightarrow$$3d$    & $-0.027\pm0.015$       & 0.214$\pm$0.015    & \cite{Batty:1979zr}\\
      Al         &$4f$$\rightarrow$$3d$    & $-0.760\pm0.050$       & 0.490$\pm$0.160    & \cite{Batty:1979zr}\\
      Si         &$4f$$\rightarrow$$3d$    & $-0.130\pm0.015$       & 0.800$\pm$0.033    & \cite{Batty:1979zr}\\
      P          &$4f$$\rightarrow$$3d$    & $-0.330\pm0.08$        & 1.440$\pm$0.120    & \cite{Backenstoss:1972yg}\\
      S          &$4f$$\rightarrow$$3d$    & $-0.550\pm0.06$        & 2.330$\pm$0.200    & \cite{Backenstoss:1972yg}\\
      Cl         &$4f$$\rightarrow$$3d$    & $-0.770\pm0.40$        & 3.80$\pm$1.0       & \cite{Backenstoss:1972yg}\\
      Co         &$5g$$\rightarrow$$4f$    & $-0.099\pm0.106$       & 0.64$\pm$0.25      & \cite{Batty:1979zr}\\
      Ni         &$5g$$\rightarrow$$4f$    & $-0.180\pm0.070$       & 0.59$\pm$0.21      & \cite{Barnes:1974iu}\\
      Cu$^1$     &$5g$$\rightarrow$$4f$    & $-0.240\pm0.220$       & 1.650$\pm$0.72     & \cite{Barnes:1974iu}\\
      Cu$^2$     &$5g$$\rightarrow$$4f$    & $-0.377\pm0.048$       & 1.35$\pm$0.17      & \cite{Batty:1979zr}\\
      Ag         &$6h$$\rightarrow$$5g$    & $-0.18 \pm0.12$        & 1.54$\pm$0.58      & \cite{Batty:1979zr}\\
      Cd         &$6h$$\rightarrow$$5g$    & $-0.40 \pm0.10$        & 2.01$\pm$0.44      & \cite{Batty:1979zr}\\
      In         &$6h$$\rightarrow$$5g$    & $-0.53 \pm0.15$        & 2.38$\pm$0.57      & \cite{Batty:1979zr}\\
      Sn         &$6h$$\rightarrow$$5g$    & $-0.41 \pm0.18$        & 3.18$\pm$0.64      & \cite{Batty:1979zr}\\
    \end{tabular*}
  \end{ruledtabular}
\end{table}

\section{Result}
\label{sec:4}
In this section, we show numerical results calculated by using the approach shown in the previous section. First we show the optical potential parameters $V_0$ and $W_0$ in various kaonic atoms. Then, we confirm their universality.

\subsection{Determining potential parameters}
Let us first determine the optical potential parameters $V_0$ and $W_0$ so as to reproduce the central value of each datum shown in Table \ref{tab:Katomdata}. These data are taken from Refs.~\cite{Backenstoss:1972yg,Batty:1979zr,Barnes:1974iu}.

We calculate the Klein-Gordon equation (\ref{KG}) with optical potential (\ref{strong}) and the Coulomb potential (\ref{Coulomb}) for various kaonic atoms using the following values of the hadron masses,
\begin{align}
  m_{K^-} & = 493.677\ \mathrm{MeV}, \\
  m_N     & = 938.919\ \mathrm{MeV}.
\end{align}
Here we have taken the isospin average of the nucleon masses.
We use a kaonic atom reduced mass given by
\begin{align}
  \mu & = A\ m_N m_{K^-}/(A\ m_N + m_{K^-}),
\end{align}
with mass number $A$. Here we ignore an effect of the nuclear binding energy. We have confirmed that the effect is negligibly small.

Let us focus on a Cu kaonic atom as an example. In the Cu kaonic atom, $4f$ orbit is the last orbit. {The strong interaction with $5g$ orbit may be ignored and the level shift and absorption width are extracted from the transition $5g \rightarrow 4f$.} There are two data for the Cu kaonic atom. We use a Cu$^1$ datum shown in Table~\ref{tab:Katomdata} for this example. We search the potential parameters which reproduce the shift and width, $-0.240 \mathrm{keV}$ and $1.650 \mathrm{keV}$ in the range of $0~\mathrm{MeV}\leq V_0,\ W_0 \leq 300~\mathrm{MeV}$. We obtain three potential parameters within this range. The result is shown in Table \ref{tab:pot_Cu}. These potentials provide the same repulsive shift and width, but have different features.  The first potential has a larger imaginary part than its real part. The second has a larger real part than its imaginary part. The third has the largest real part of three and a small imaginary part.

\begin{table}
  \caption{Obtained potential parameters for Cu$^1$ kaonic atom.}
  \label{tab:pot_Cu}
  \begin{ruledtabular}
    \begin{tabular*}{8.6cm}{@{\extracolsep{\fill}}cccc}
      Used datum   &Potential type & $V_0$[MeV]   & $W_0$[MeV]  \\ \hline
      Cu$^1$       &1    & 79.5          & 114.5    \\
                   &2    & 78.0          & 20.0     \\
                   &3    & 199.5         & 28.0     \\
    \end{tabular*}
  \end{ruledtabular}
\end{table}

\begin{figure}[t]
  \centering
  \includegraphics[width=0.45\textwidth]{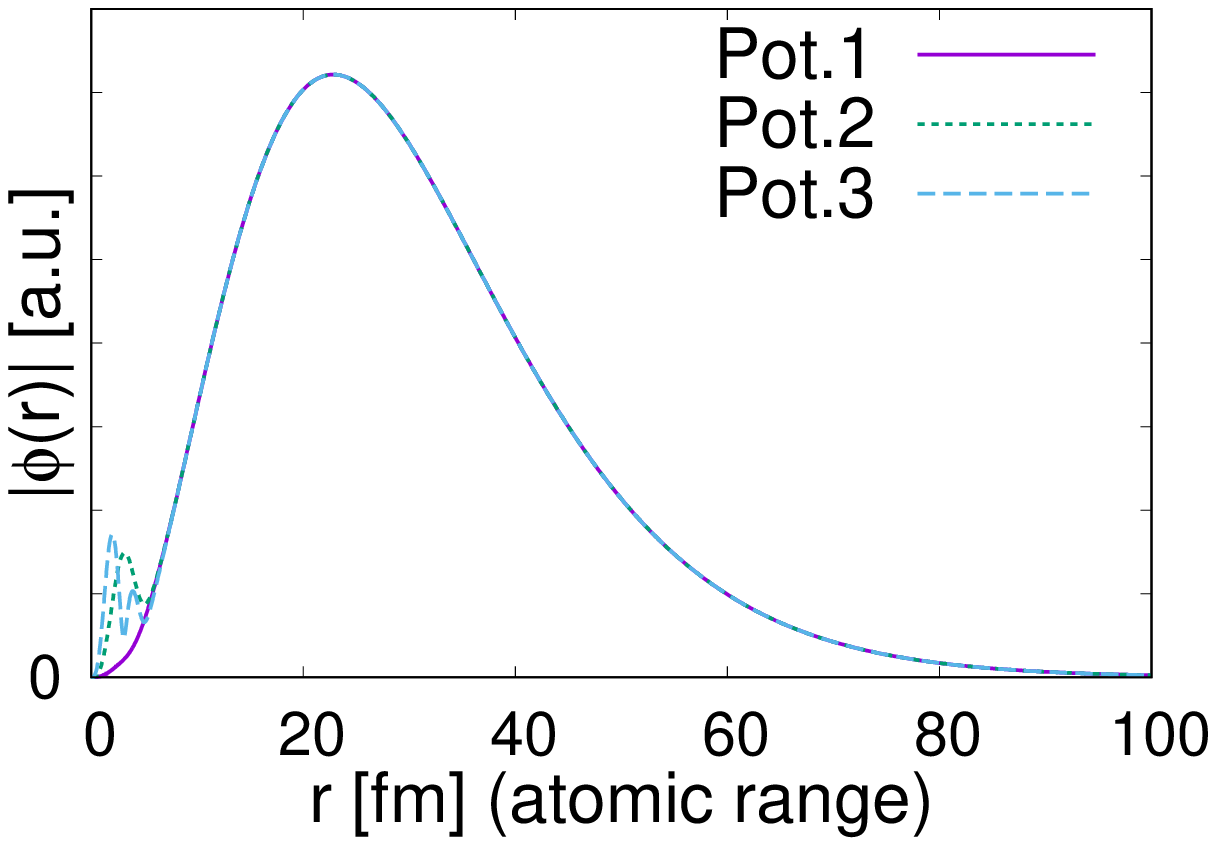}
  \caption{ Wavefunctions squared of the atomic states with $\ell = 3$ (last orbit) calculated with Potentials 1, 2 and 3 for the Cu$^1$ kaonic atom in the atomic radial range.}
  \label{fig:wf}
  \includegraphics[width=0.45\textwidth]{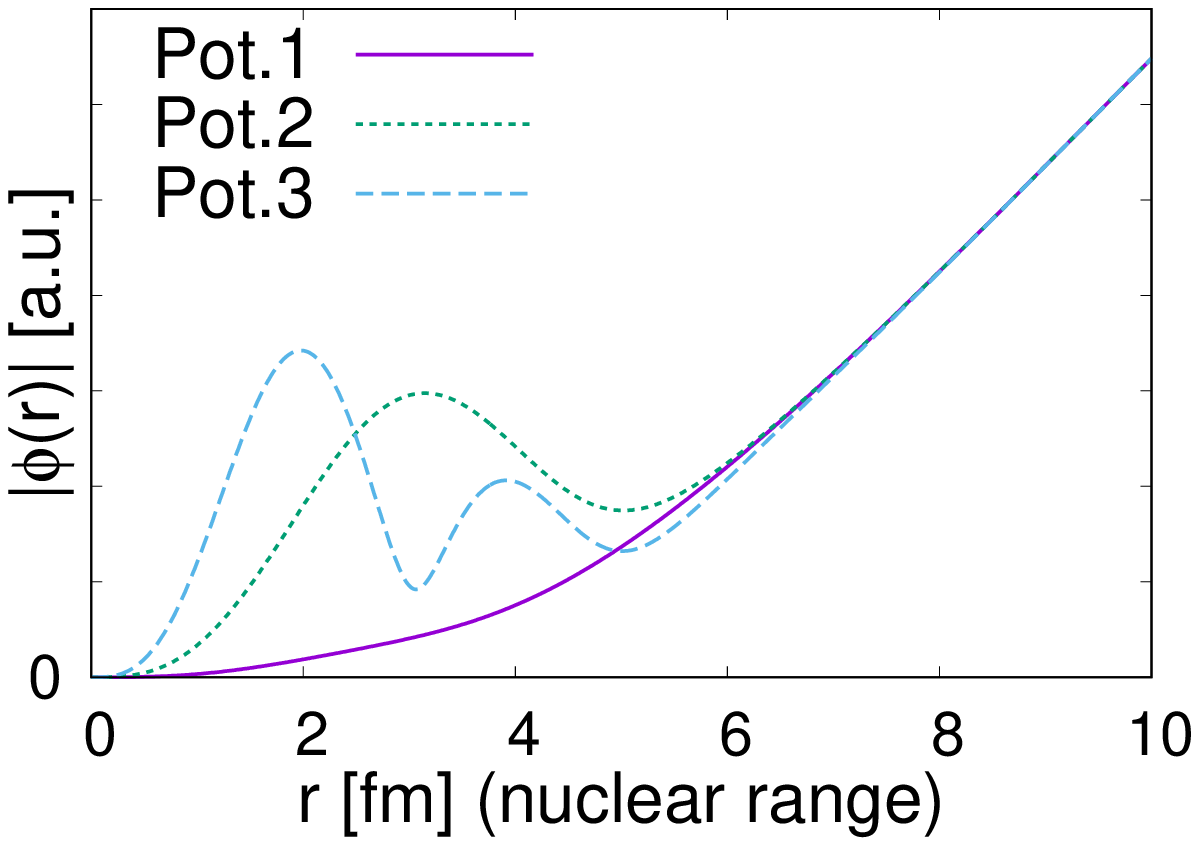}
  \caption{The same as in Fig.1, but in the nuclear radial range. }.
  \label{fig:wf2}
\end{figure}

We calculate wavefunctions using these potentials and show those of the Cu kaonic atom in Figs~\ref{fig:wf} and~\ref{fig:wf2}. Fig~\ref{fig:wf} shows wavefunctions of the Cu kaonic atom in an atomic scale ($r=0\sim100$ fm) while Fig~\ref{fig:wf2} shows wavefunctions of the kaonic atom in a nuclear scale ($r=0\sim10~\mathrm{fm}$).
These potentials provide similar wavefunctions in the atomic scale, while the number of nodes which corresponds to the number of the nuclear bound states, is different from each other in the nuclear scale. The wavefunction with Potential~1 has no nodes, the wavefunction with Potential~2 has one node and the wavefunction with Potential~3 has two nodes.

\begin{figure}[t]
  \includegraphics[width=0.45\textwidth]{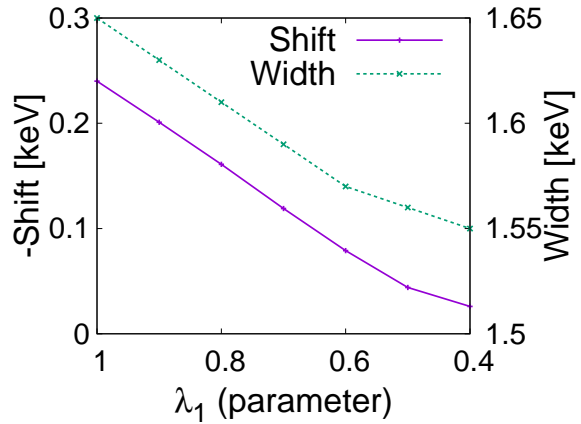}
  \caption{Energy shifts and absorption widths of the atomic state with $\ell =3$ calculated for the Cu$^1$ kaonic atom by Potential~1 as a function of $\lambda_1$, which is introduced to control the strength of the imaginary part. The solid (dotted) line denoting the shift (width) uses the left (right) axis. }
  \label{fig:lambda1}
\end{figure}

The Cu kaonic atom with Potential~1 has no nuclear states with the same angular momentum as the atomic state with $\ell=3$ in which we are interested here. Thus, the nuclear states are not responsible for the repulsive shift in Potential~1. In order to see the role of the imaginary part of Potential~1 for the repulsive shift, we reduce the strength of the imaginary part of the potential by introducing a factor $\lambda_1$ as $\Im V_{\rm opt}(r) \rightarrow \lambda_1 \Im V_{\rm opt}(r)$ and changing
$\lambda_1$ from 1 to 0.4 without changing the real part. The energy shift and the width for each $\lambda_1$ is plotted in Fig.~\ref{fig:lambda1}. This figure shows that, as the strength of the imaginary part decreases, the energy shift does also decrease. This implies that the large imaginary part is responsible for the repulsive shift in Potential~1.

Potentials 2 and 3 provide nuclear states with $\ell=3$ below the atomic $4f$ state. We find one nuclear state with $\ell=3$ for Potential~2 at $-E_B - i \Gamma/2 = (-5.8 - i 22.4/2)\ \mathrm{MeV}$ and two nuclear states for Potential~3 at $-E_B - i \Gamma/2 = (-102.5-i57.0/2)\ \mathrm{MeV}$ as the ground state and at $-E_B - i \Gamma/2 = (-17.7-i32.0/2)\ \mathrm{MeV}$ as a radial excited state. The detailed values of these spectra are not important. What we emphasize here is the fact that a shallow nuclear state exists with the same quantum number to the atomic state, and this nuclear state definitely couples to the atomic state and repels it up due to the level repulsion. In order to confirm this speculation, we introduce a parameter $\lambda_2$ in the optical potential as $V_{\rm opt}(r) \rightarrow \lambda_2 V_{\rm opt}(r)$ to change the position of the nuclear state. Changing $\lambda_2$ from 1.0 to 0.6, we find that the binding energy of the nuclear state gets smaller and negative for $\lambda_2 {\le} 0.8$. (The state with a negative binding energy is a resonance thanks to the centrifugal barrier with $\ell=3$.) The nuclear state with the negative binding energy is located above the atomic state. In such a case, the energy shift of the atomic state gets attractive. This is because the atomic state lies below the nuclear state and the level repulsion pushes down the atomic state. We see this situation with $\lambda_2 = 0.6, 0.7, 0.8$ in Table~\ref{tab:lambda2}. In this way, we conclude that in Potential~2 and 3 the nuclear state is responsible for the repulsive shift of the atomic state. 
The reason that the potentials 2 and 3 have a smaller imaginary part than Potential~1 is as follows. The absorption width may be given by the overlap of the wavefunction of the atomic state and the imaginary part of the potential as
\begin{align}
  \Gamma \sim \bra{\mathrm{Atom}}\Im V_{\mathrm{opt}}\ket{\mathrm{Atom}}.
\end{align}
As seen in Fig.~\ref{fig:wf2}, thanks to the nuclear states, the wavefunctions for Potentials 2 and 3 have a larger contribution than that of Potential~1 in the nuclear length scale. Because the optical potential has values in the nuclear range, the overlap integrals get sufficiently large even with small imaginary parts of the optical potential to reproduce the observed width.

\begin{figure}[t]
  \includegraphics[width=0.45\textwidth]{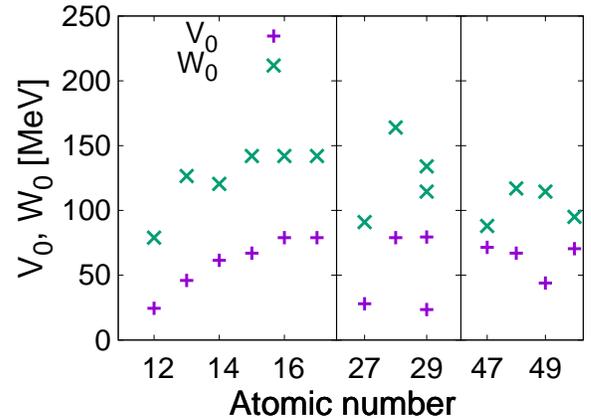}
  \caption{Atomic dependence of the potential parameters for Potential~1. The plus ($+$) and cross ($\times$) signs denote the parameters for the real and imaginary parts, respectively.}
  \label{fig:Pot1}
\end{figure}

\begin{figure}[t]
  \includegraphics[width=0.45\textwidth]{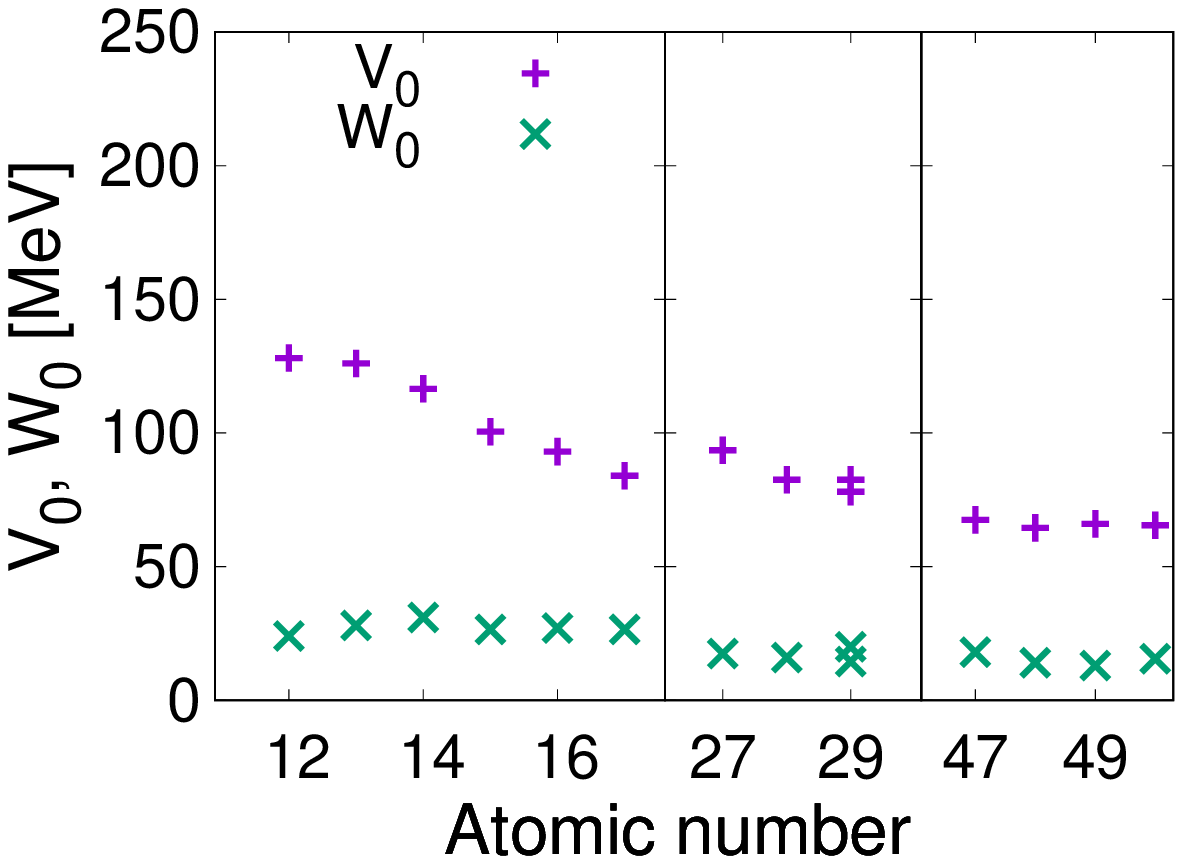}
  \caption{The same as in Fig.4, but for Potential~2.}
  \label{fig:Pot2}
\end{figure}

\begin{figure}[t]
  \includegraphics[width=0.45\textwidth]{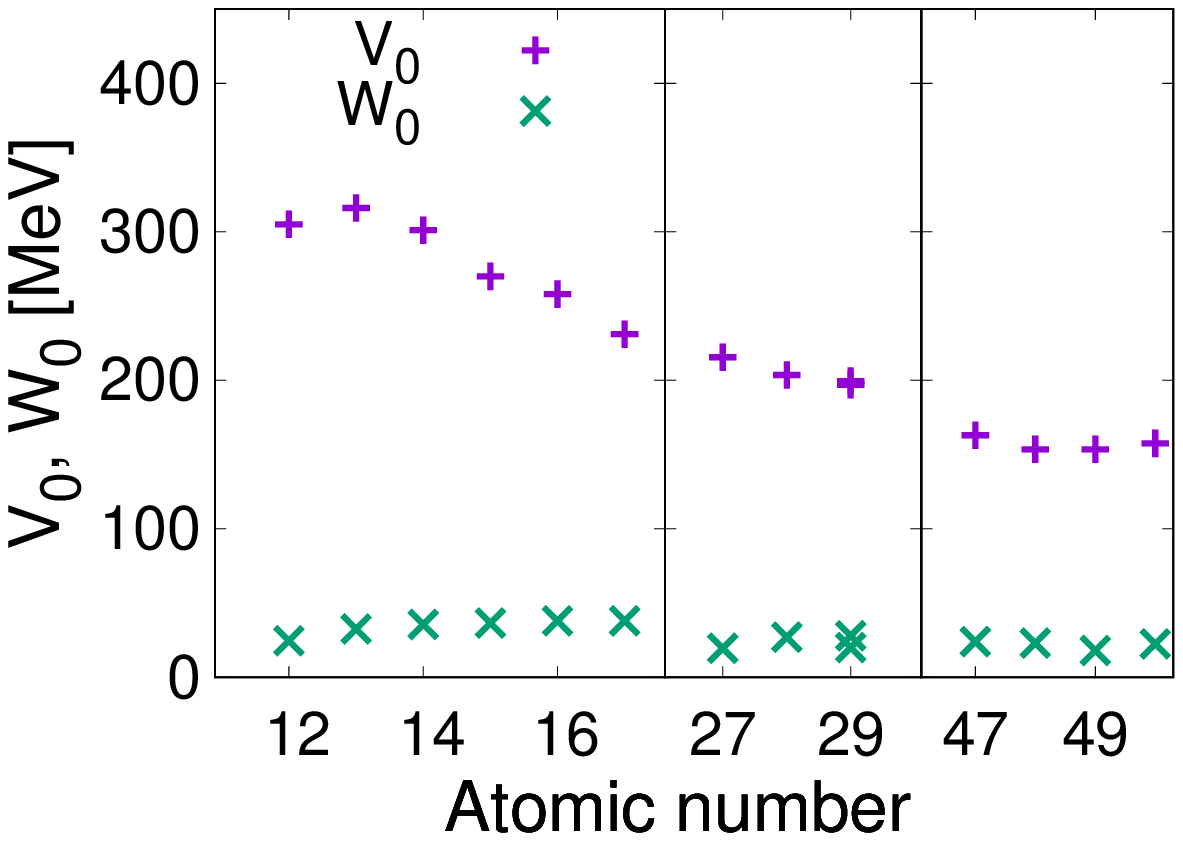}
  \caption{The same as in Fig.4, but for Potential~3.}
  \label{fig:Pot3}
\end{figure}

Now let us see the potential parameters obtained by other kaonic atoms. We find a very similar trend to the Cu case, that is, the potentials are classified into three categories; Potential~1 having a large imaginary part, Potential~2 having a small imaginary part and Potential~3 having a deep real part and a small imaginary part. Potential~1 provides no nuclear states with the same quantum numbers to the atomic state of interest, while Potentials 2 and 3 have nuclear states. We show the numerical results for the obtained potential parameters in Tables~\ref{tab:potential1}, \ref{tab:potential2} and \ref{tab:potential3} for Potentials 1, 2 and 3, respectively. In Figs.~\ref{fig:Pot1}, \ref{fig:Pot2} and \ref{fig:Pot3}, we plot the values of the potential parameters for each kaonic atom to see the atomic number dependence. One finds in Fig.~\ref{fig:Pot1} that the imaginary parts of Potential~1 are larger than those of the real parts independently to the atomic number, and that the values themselves are scattered. For Potentials 2 and 3, Figs.~\ref{fig:Pot2} and \ref{fig:Pot3} show that the real parts of the potentials are larger than their imaginary parts again independently to the atomic number, and that the real parts tend to decrease as the atomic number increases.

\begin{table}
  \caption{Energy shifts and absorption widths of the atomic and nuclear states with $\ell =3$ calculated for the Cu$^1$ kaonic atom by Potential~2 as a function of $\lambda_2$, which is introduced to control the strength of the potential. We the nuclear states get resonances, the energy shifts of the atomic states becomes attractive. }
  \label{tab:lambda2}
  \begin{ruledtabular}
    \begin{tabular*}{8.6cm}{@{\extracolsep{\fill}}c|cc|cc}
      $\lambda _2$   &\multicolumn{2}{c|}{kaonic atom}   &\multicolumn{2}{c}{kaonic nucleus}\\ \hline
      &Shift [keV]  &Width [keV]  &$E_B$ [MeV] &Width [MeV]    \\
      1.0           &0.240     &1.65   &5.8    &22.4\\
      0.9           &0.168     &2.18   &1.8    &17.5\\
      0.8           &-0.216    &2.57   &-2.0   &13.3\\
      0.7           &-0.648    &2.05   &-5.3   &8.2\\
      0.6           &-0.696    &1.19   &-8.0   &2.3\\
    \end{tabular*}
  \end{ruledtabular}
\end{table}

\begin{table}
  \caption{Determined potential parameters of Potential~1 for each kaonic atom.}
  \label{tab:potential1}
  \begin{ruledtabular}
    \begin{tabular*}{8.6cm}{@{\extracolsep{\fill}}ccc}
      Used datum.   & $V_0$[MeV]& $W_0$[MeV]  \\ \hline
      Mg           & 24.5       & 79.0     \\
      Al           & 46.0       & 126.5    \\
      Si           & 61.5       & 120.5    \\
      P            & 67.0       & 142.0    \\
      S            & 79.0       & 142.0    \\
      Cl           & 79.0       & 142.0    \\ \hline
      Co           & 28.0       & 91.0     \\
      Ni           & 79.0       & 164.0    \\
      Cu$^1$       & 79.5       & 114.5    \\
      Cu$^2$       & 23.5       & 134.0    \\ \hline
      Ag           & 71.5       & 88.0     \\
      Cd           & 67.5       & 117.0    \\
      In           & 44.0       & 114.5    \\
      Sn           & 70.5       & 95.0     \\
    \end{tabular*}
  \end{ruledtabular}
\end{table}

\begin{table}
  \caption{The same as in Table 5, but of Potential~2.}
  \label{tab:potential2}
  \begin{ruledtabular}
    \begin{tabular*}{8.6cm}{@{\extracolsep{\fill}}ccc}
      Used datum.    & $V_0$[MeV]   & $W_0$[MeV]  \\ \hline
      Mg           & 128.0      & 24.0     \\
      Al           & 126.0      & 28.0     \\
      Si           & 116.5      & 31.0     \\
      P            & 100.5      & 26.5     \\
      S            & 93.0       & 27.0     \\
      Cl           & 84.0       & 26.5     \\ \hline
      Co           & 93.5       & 17.5     \\
      Ni           & 82.5       & 16.0     \\
      Cu$^1$       & 78.0       & 20.0     \\
      Cu$^2$       & 82.5       & 14.5     \\ \hline
      Ag           & 67.5       & 18.0     \\
      Cd           & 64.5       & 14.0     \\
      In           & 66.0       & 13.0     \\
      Sn           & 65.5       & 15.5     \\
    \end{tabular*}
  \end{ruledtabular}
\end{table}
\begin{table}
  \caption{The same as in Table 5, but of Potential~3.}
  \label{tab:potential3}
  \begin{ruledtabular}
    \begin{tabular*}{8.6cm}{@{\extracolsep{\fill}}cccc}
      Used datum.    & $V_0$[MeV]   & $W_0$[MeV]  \\ \hline
      Mg           & 305.0      & 24.5     \\
      Al           & 316.0      & 32.5     \\
      Si           & 301.0      & 35.5     \\
      P            & 270.0      & 36.5     \\
      S            & 258.0      & 38.0     \\
      Cl           & 231.0      & 38.0     \\ \hline
      Co           & 215.5      & 19.5     \\
      Ni           & 203.5      & 27.0     \\
      Cu$^1$       & 199.5      & 28.0     \\
      Cu$^2$       & 197.0      & 20.0     \\ \hline
      Ag           & 163.0      & 24.0     \\
      Cd           & 153.5      & 23.0     \\
      In           & 153.5      & 18.0     \\
      Sn           & 157.5      & 22.5     \\
    \end{tabular*}
  \end{ruledtabular}
\end{table}

In summary, we find that the observed data of the kaonic atoms suggest two types of the optical potentials; one potential has a large imaginary part without providing nuclear states with the same quantum number to the atomic state (Potential~1), and the other (Potentials 2 and 3) has a sufficiently large real part to produce a shallow nuclear bound state. For the origin of the repulsive shift of the kaonic atom, in the former potential, the imaginary part works repulsively to the atomic state, while in the latter potential the nuclear bound state repels up the atomic state.

\subsection{Confirmation of  universality}
In the previous section, for each kaonic atom, two scenarios are possible to explain the repulsive shifts of the atomic states. Here we pin down which scenario is realized in the observed kaonic atom by checking the universality of the optical potential obtained by one kaonic atom datum for other kaonic atoms. We calculate the energy shifts and the absorption widths for various kaonic atoms with the optical potential obtained in the previous section.

\begin{figure}[t]
  \includegraphics[width=0.45\textwidth]{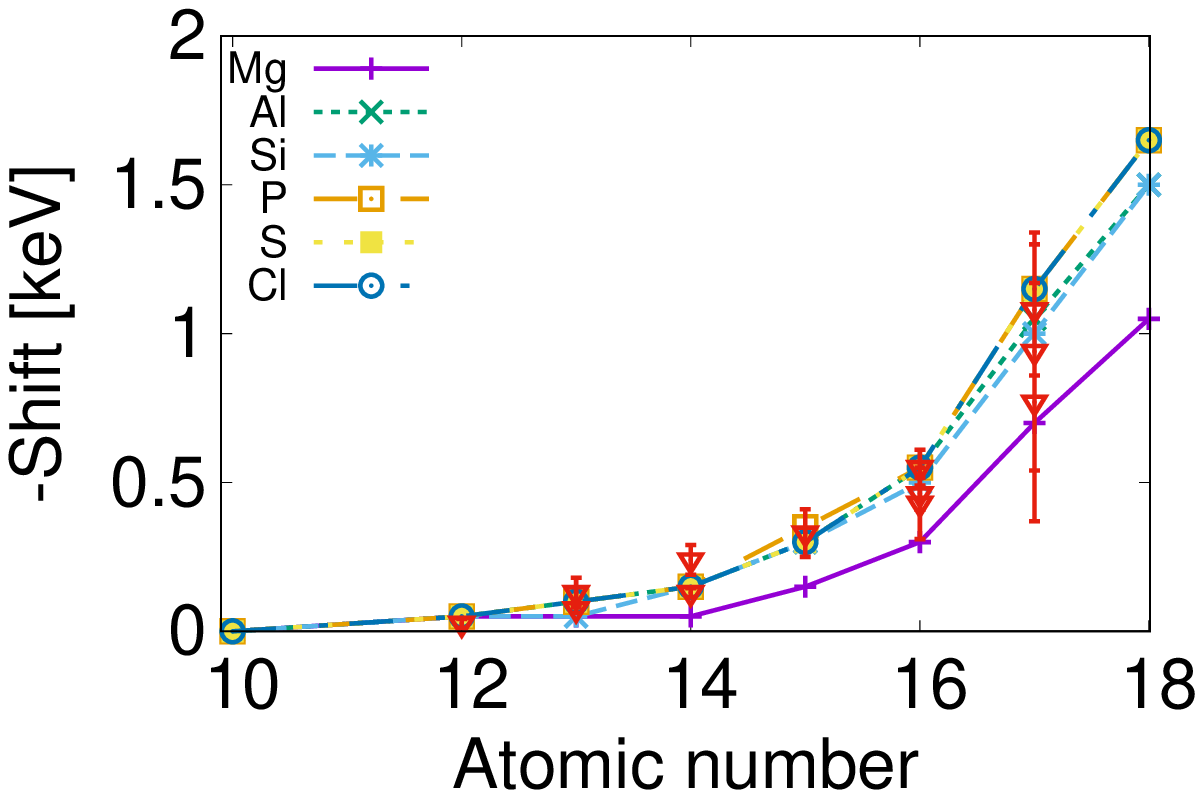}
  \includegraphics[width=0.45\textwidth]{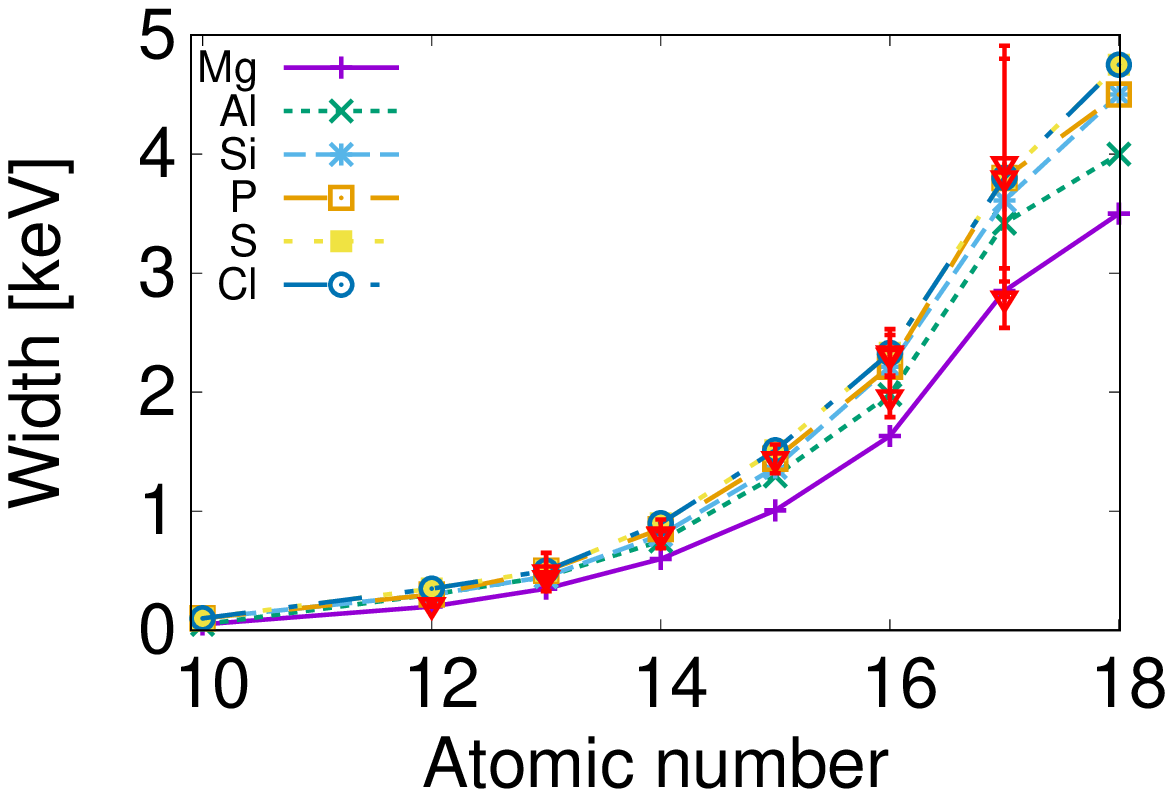}
  \caption{Energy shifts (upper plot) and absorption width (lower plot) of the kaonic atoms with $\ell = 2$ calculated with Potential~1. Each line corresponds to the results obtained by the potential parameters determined by the datum of the indicated atom. The points with error bars are the experimental data.}
  \label{fig:3d_pot1}
\end{figure}

\begin{figure}[t]
  \includegraphics[width=0.45\textwidth]{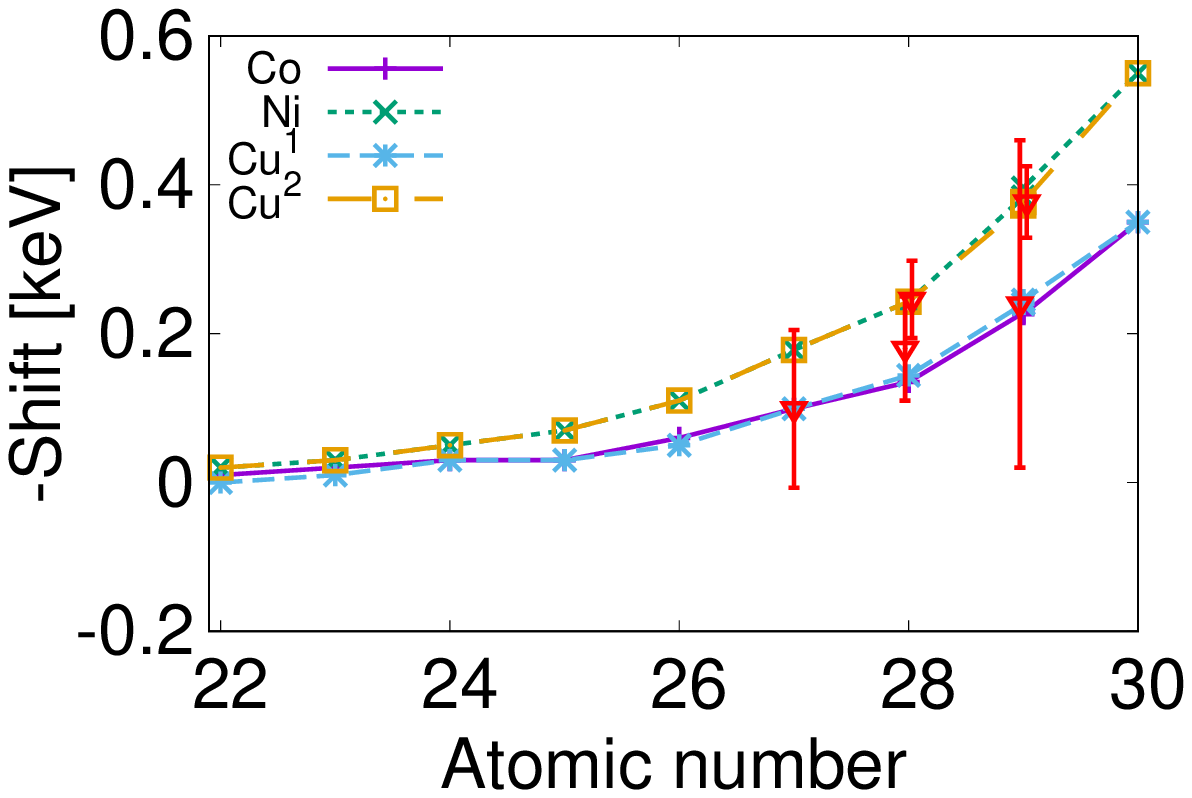}
  \includegraphics[width=0.45\textwidth]{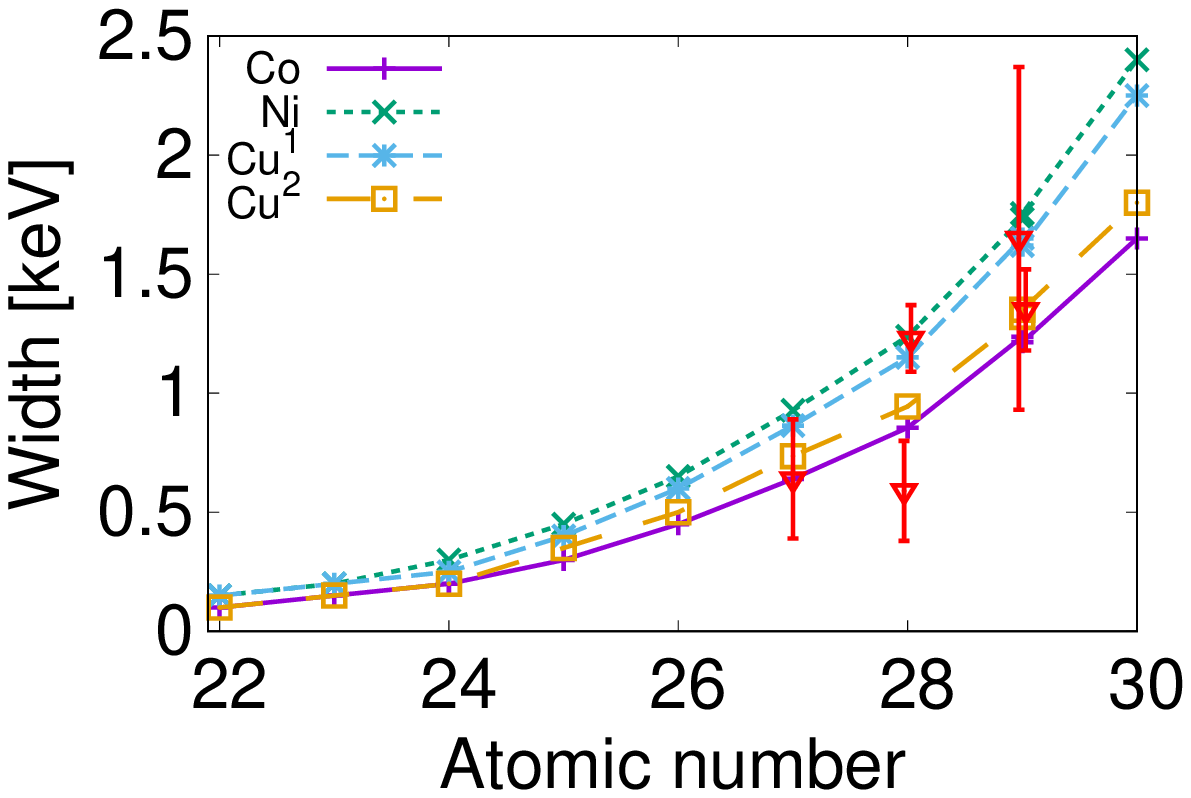}
  \caption{The same in Fig.~\ref{fig:3d_pot1}, but for the kaonic atoms with $\ell = 3$}
  \label{fig:4f_pot1}
\end{figure}

\begin{figure}[t]
  \includegraphics[width=0.45\textwidth]{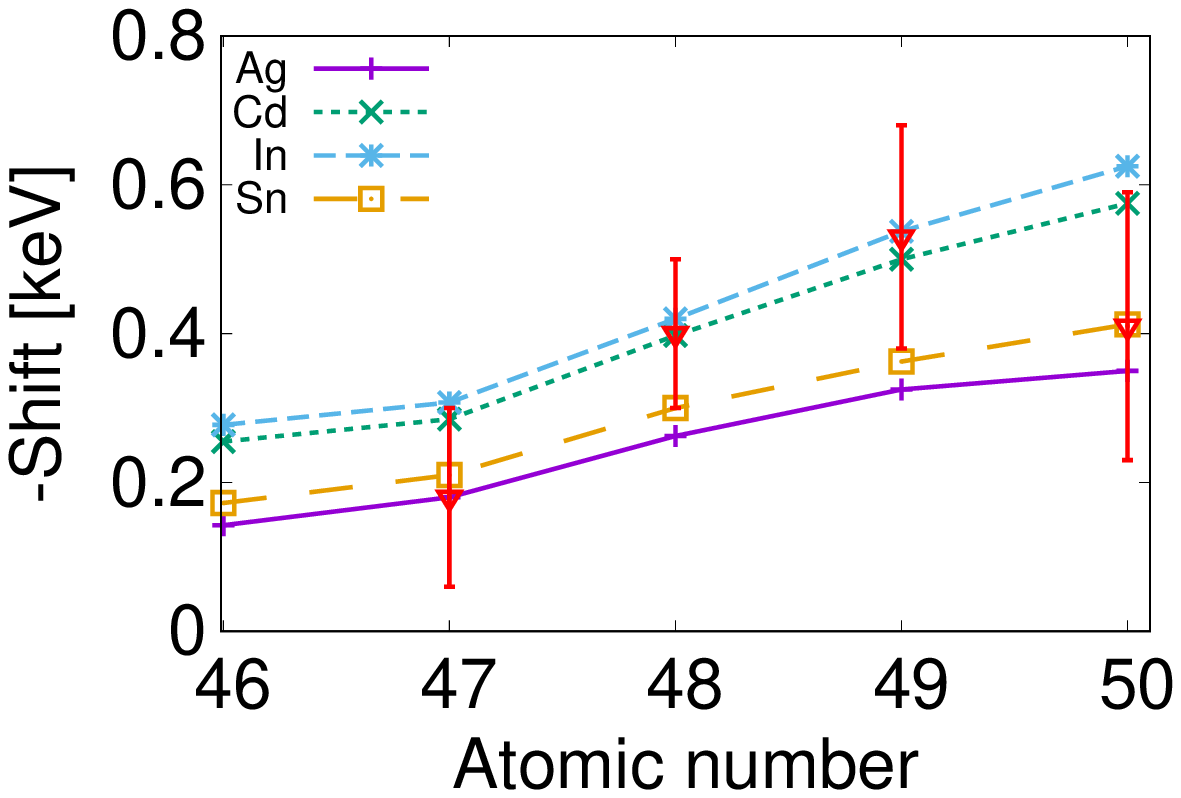}
  \includegraphics[width=0.45\textwidth]{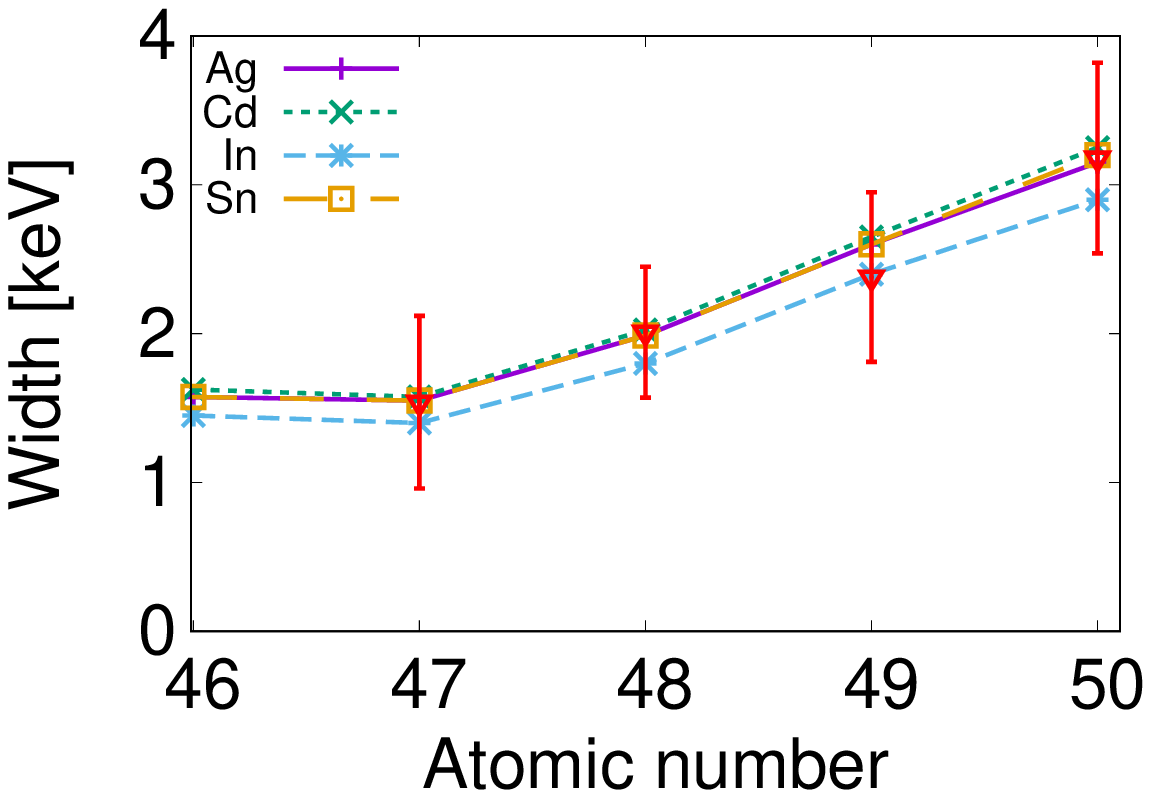}
  \caption{The same in Fig.~\ref{fig:3d_pot1}, but for the kaonic atoms with $\ell = 4$.}
  \label{fig:5g_pot1}
\end{figure}

\begin{figure}[t]
  \includegraphics[width=0.45\textwidth]{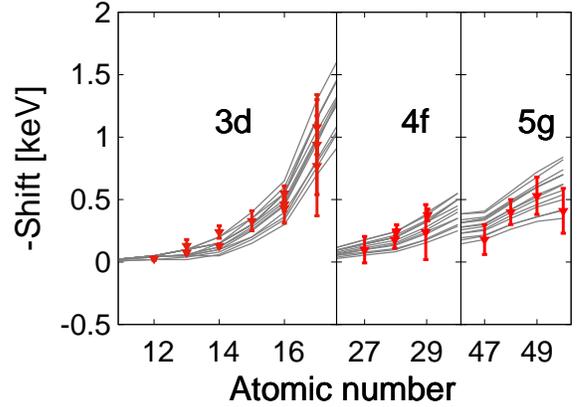}
  \caption{Summary of energy shifts of kaonic atoms in the last orbit calculated with Potential~1. The values are same as Figs.~\ref{fig:3d_pot1}, \ref{fig:4f_pot1} and \ref{fig:5g_pot1}.}
  \label{fig:pot1_all}
\end{figure}

We show in Fig.~\ref{fig:3d_pot1} the energy shifts and the absorption widths of the $3d$ states of kaonic atoms calculated with Potential~1 in comparison with experimental data. Each line is calculated with a set of potential parameters $(W_0, V_0)$ fitted by the data of a kaonic atom. This figure shows that all of the lines are with the errors of the experimental data, and we find that the calculation with Potential~1 is consistent with all of the observed experiments for the $3d$ states. In the same way, we plot the energy shifts and the absorption widths for the $4f$ states in Fig.~\ref{fig:4f_pot1} and for the $5g$ states in Fig.~\ref{fig:5g_pot1}. In Fig.~\ref{fig:pot1_all} we show the shifts calculated by all of the potentials obtained in the previous section, and we find that the calculations are consistent with the experimental data. This implies that the potential with a large imaginary potential widely explains the kaonic atom data and can be the origin of the repulsive shift. This type of potential was found in Ref.~\cite{Friedman:1994hx} as $t_\mathrm{eff}\rho$.

Actually the real parts of the obtained optical potentials widely spread from 20 MeV to 80 MeV and are not determined as well as the imaginary parts. Thus, the real part of the optical potential is not sensitive to the spectra of the kaonic atoms. For precise determination of the optical potential, one should observed the nuclear states of the kaonic nucleus system.

\begin{figure}[t]
  
  \includegraphics[width=0.45\textwidth]{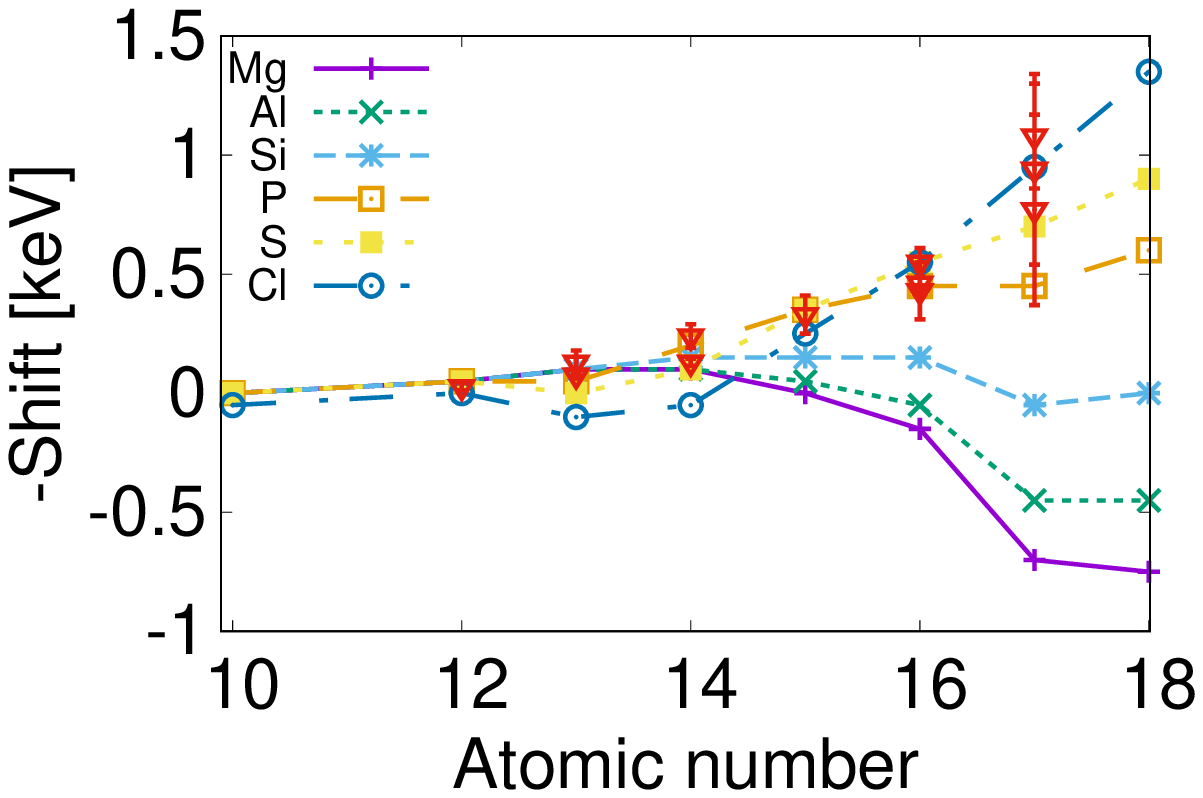}
  \includegraphics[width=0.45\textwidth]{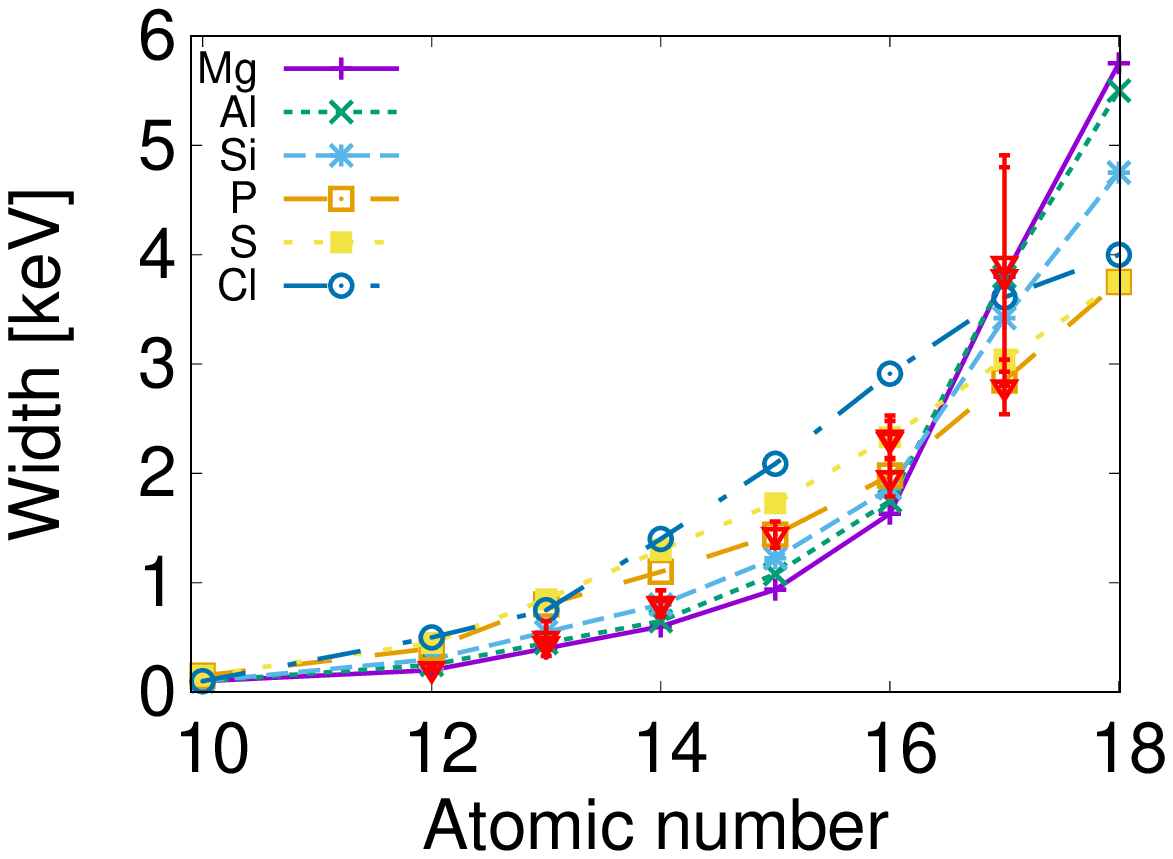}
  \caption{The same in Fig.~\ref{fig:3d_pot1}, but for Potential~2. }
  \label{fig:3d_pot2}
\end{figure}

\begin{figure}[t]
  \includegraphics[width=0.45\textwidth]{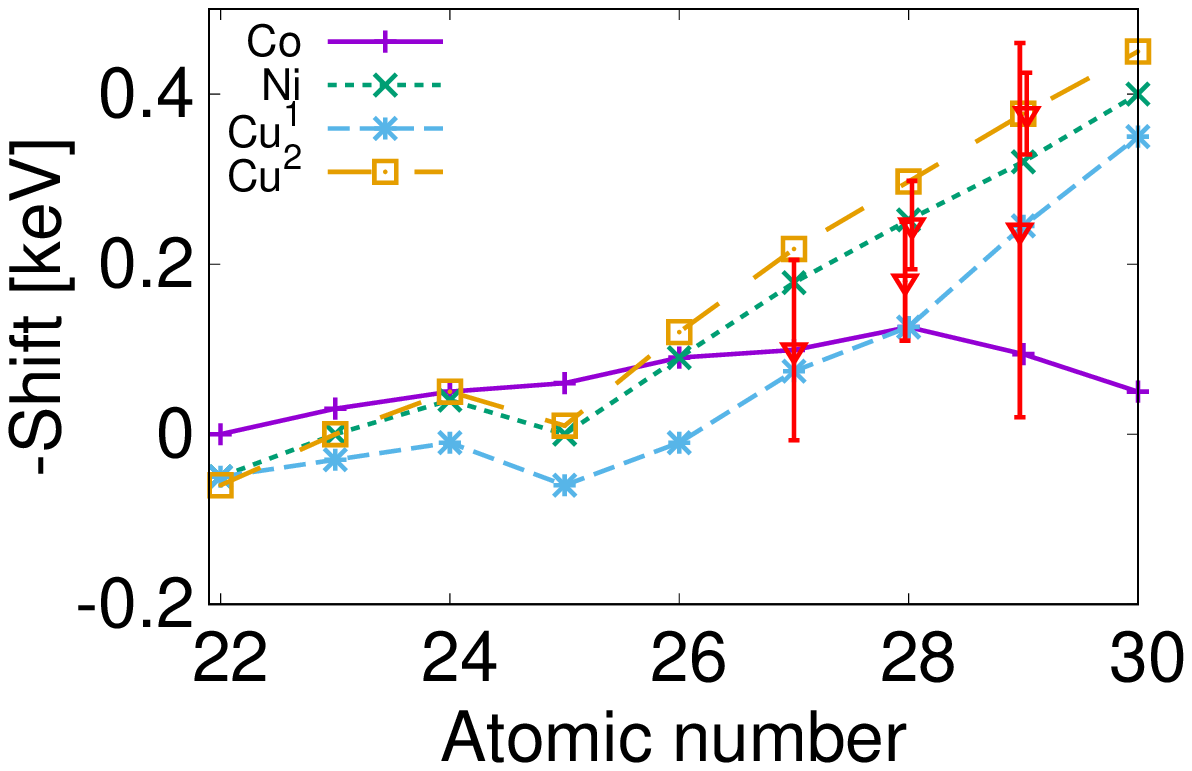}
  \includegraphics[width=0.45\textwidth]{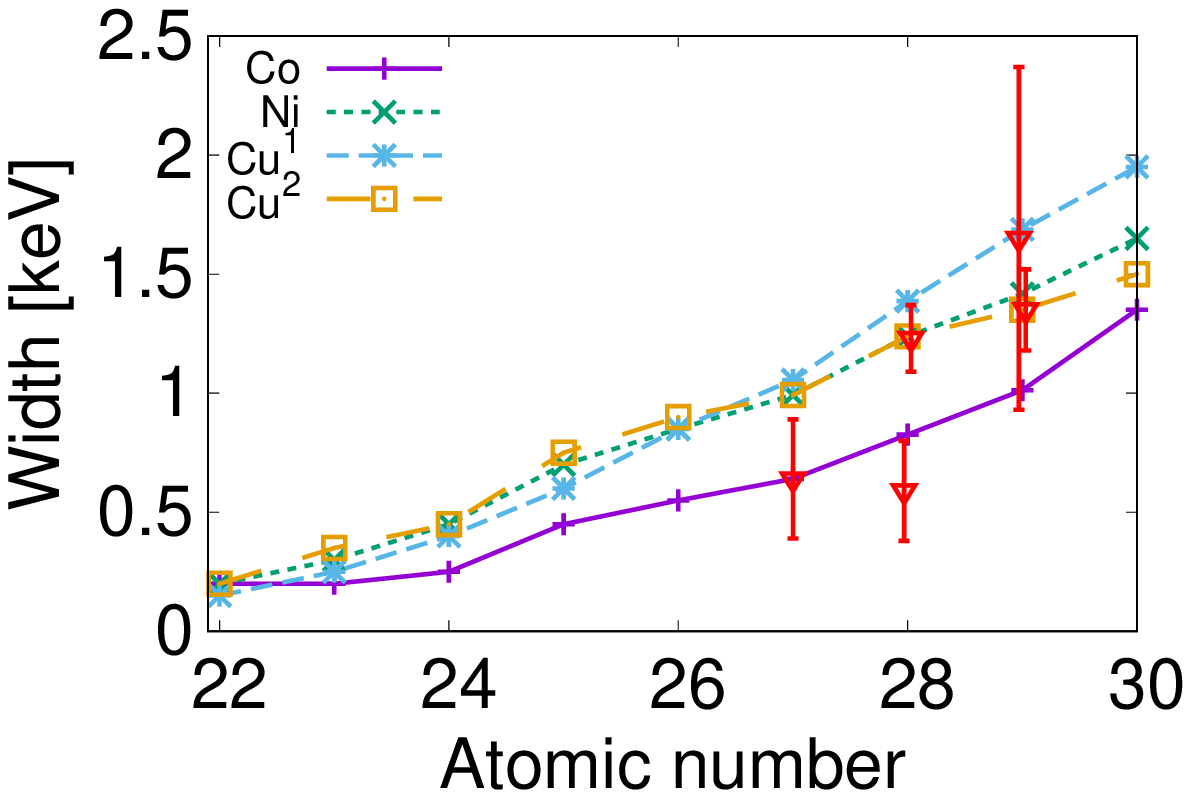}
  \caption{The same in Fig.~\ref{fig:3d_pot2}, but for the kaonic atoms with $\ell = 3$.}
  \label{fig:4f_pot2}
\end{figure}

\begin{figure}[t]
  \includegraphics[width=0.45\textwidth]{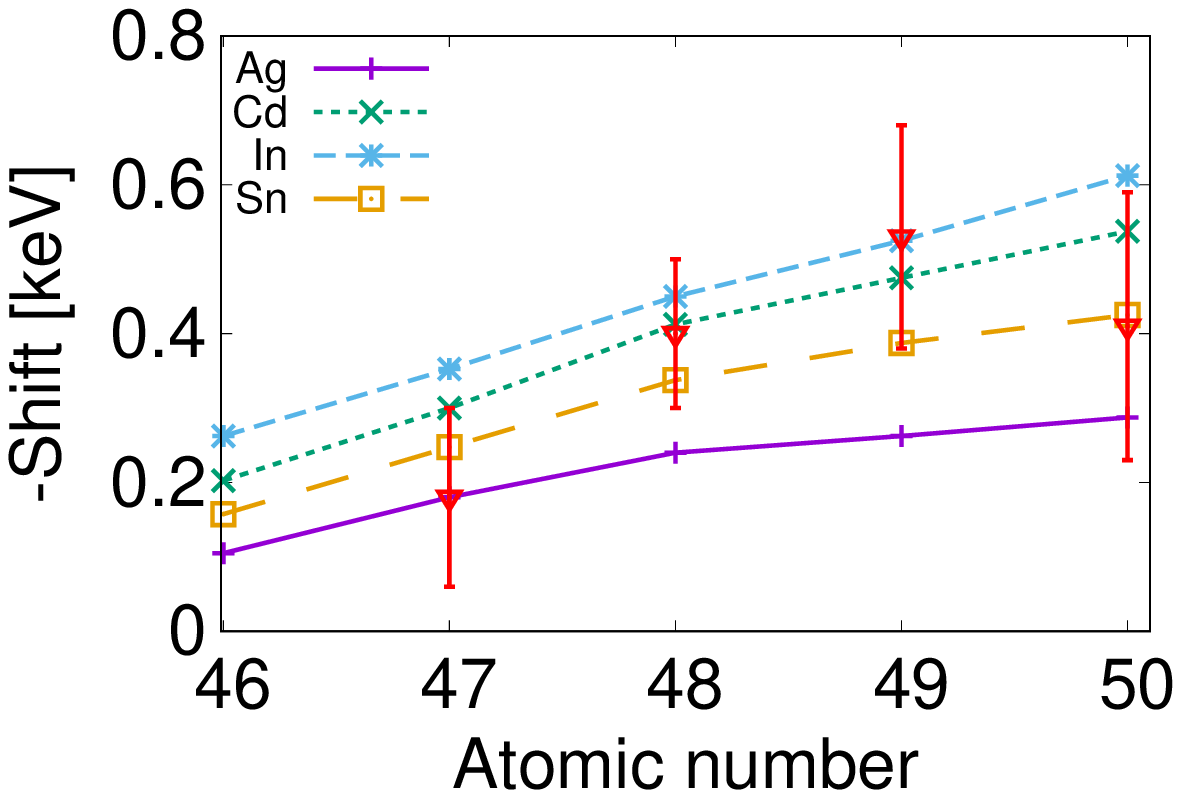}
  \includegraphics[width=0.45\textwidth]{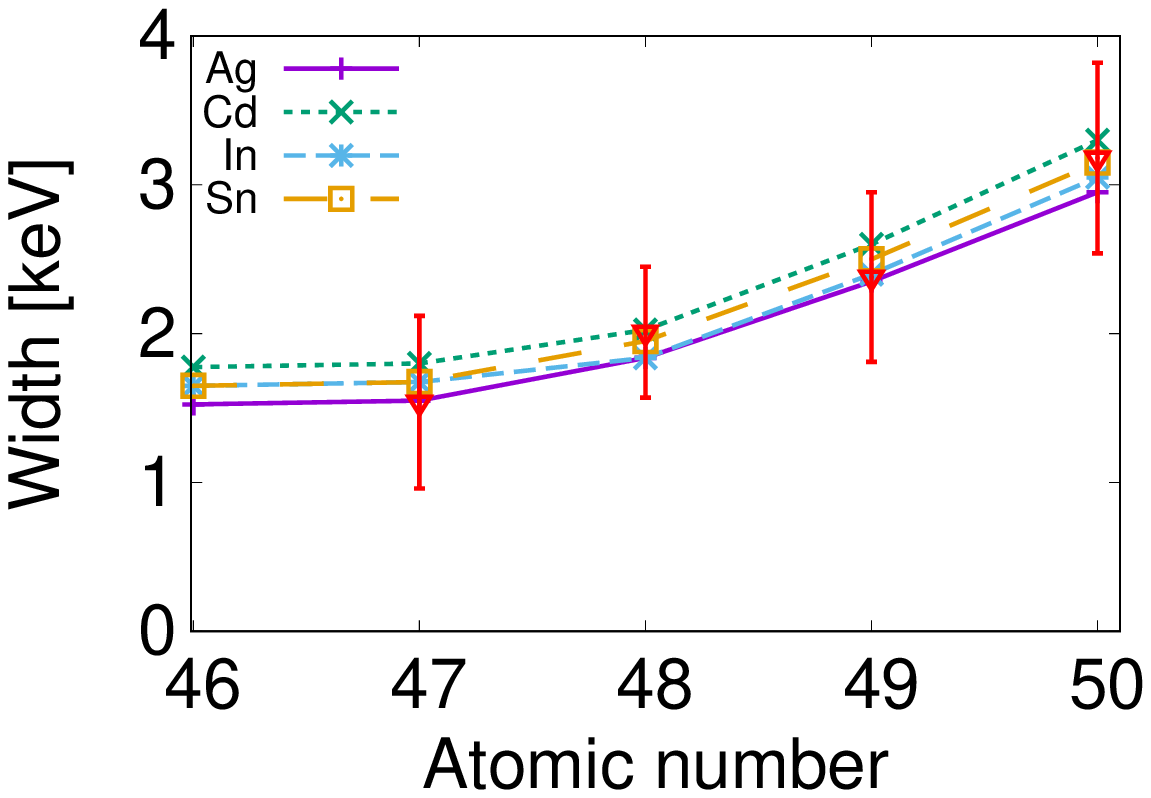}
  \caption{The same in Fig.~\ref{fig:3d_pot2}, but for the kaonic atoms with $\ell = 4$.}
  \label{fig:5g_pot2}
\end{figure}

\begin{figure}[t]
  \includegraphics[width=0.45\textwidth]{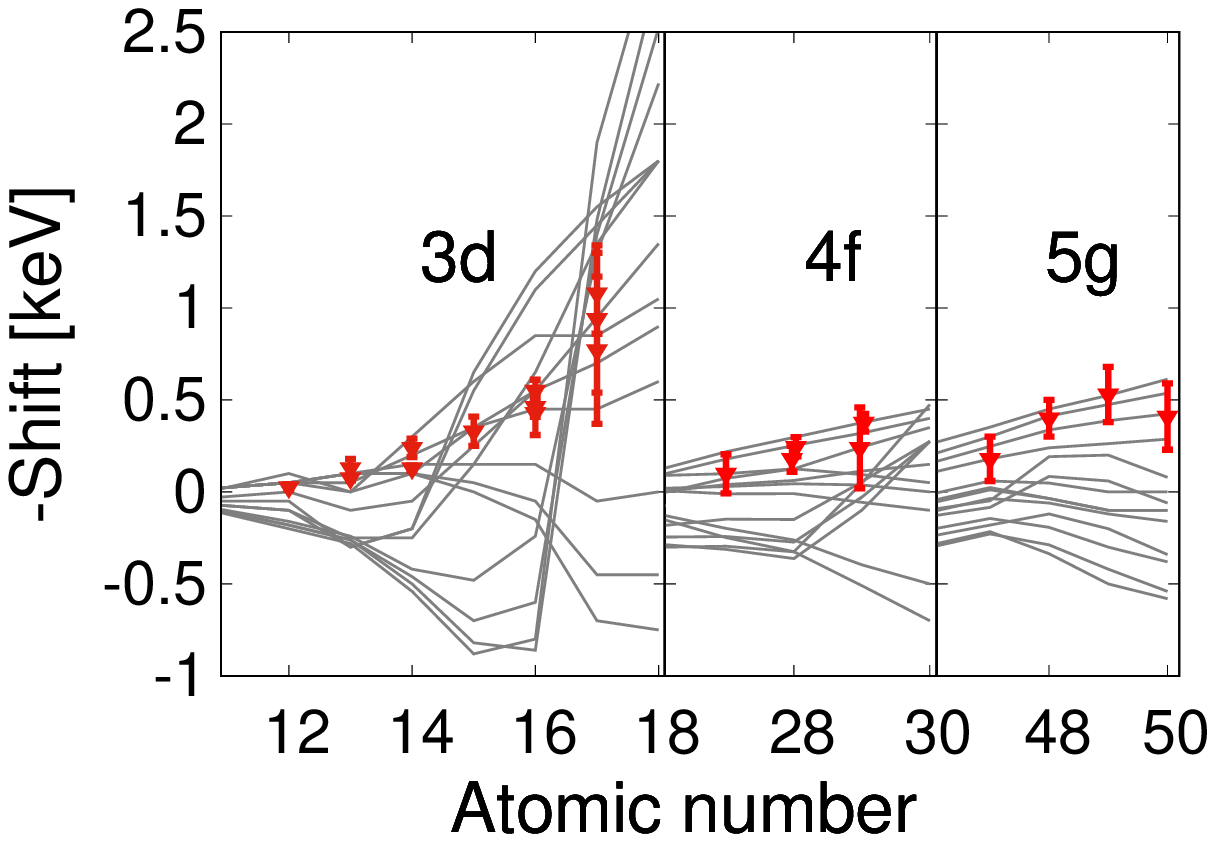}
  \caption{The same in Fig.~\ref{fig:pot1_all}, but for Potential~2.}
  \label{fig:pot2_all}
\end{figure}

\begin{figure}[t]
  \includegraphics[width=0.45\textwidth]{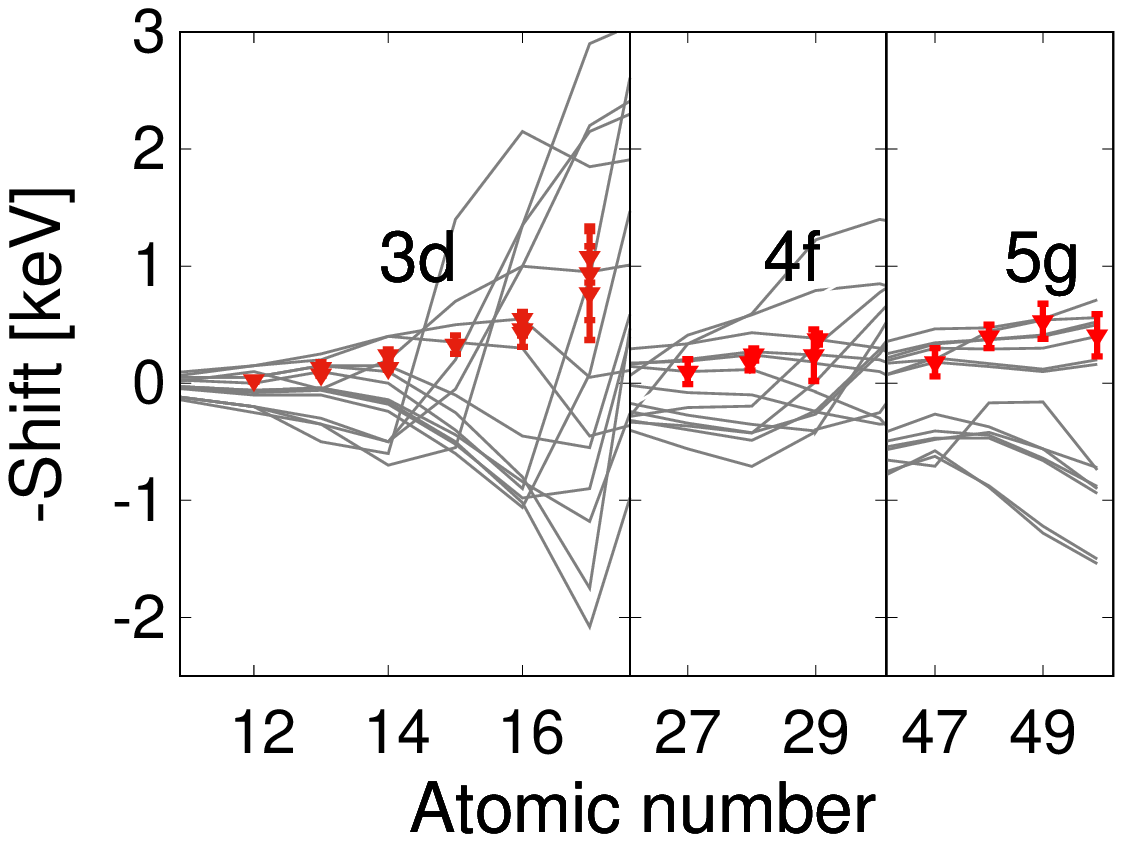}
  \caption{The same in Fig.~\ref{fig:pot1_all}, but for Potential~3.}
  \label{fig:pot3_all}
\end{figure}

The energy shifts and the absorption widths calculated with Potential~2 for the $3d$, $4f$ and $5g$ states of kaonic atoms are shown in Figs.~\ref{fig:3d_pot2}, \ref{fig:4f_pot2} and \ref{fig:5g_pot2}, respectively, and in Fig.~\ref{fig:pot2_all}, we show all of the calculation in one figure. These figures show that Potential~2 does not universally reproduce the observed shifts and widths, and Potential~2 explains a specific kaonic nucleus only. Therefore, Potential~2 cannot be a solution to explain all of the energy shifts and widths of the observed kaonic atoms, and the presence of the nuclear states is not responsible for the repulsive shift. For Potential 3, we find a discrepancy to the experiments in a similar way to Potential 2 as seen in the summary plot of Potential 3 given in Fig.~\ref{fig:pot3_all}.

The reason that Potentials~2 and 3 do not explain the repulsive shifts is as follows. If the level repulsion would be the origin of the repulsive shift, the atomic state should be sensitive to the position of the nuclear states because the mixing of these states determines the energy of the atomic state. Now each kaonic atom shares a common depth of the optical potential, but the potential size depends on the nuclear radius which is proportional to $A^{1/3}$. When the potential size is wider, the bound states have a deeper energy, which is an exercise of quantum mechanics for the bound states in a square-well potential. If the nuclear states get deeper, the effects of the level repulsion become weaker, because the level mixing takes place when the levels are close energetically. As a consequence, the energy of the atomic state is strongly dependent on the atomic nuclide. Thus, this kind of potential cannot be universally applied for the kaonic atoms. Actually, this type of potential was not found in the global fit performed in Ref.~\cite{Friedman:1994hx}.

%
%
%
%

\section{other types of potential}
\label{sec:5}
In this section, we discuss two other types of the potential in order to see the effect of the nuclear surface. One is a non-linear nuclear density potential, and the other is a linear nuclear density  potential using other parameters of nuclear radius and diffuseness. These are accepted as optical potentials in the previous analysis \cite{Friedman:1994hx}.

\subsection{Potential with non-linear density term}
We study the optical potential of kaonic atoms with non-linear term for nuclear density $\rho_N$. We have studied the optical potential of kaonic atoms under the linear nuclear density approximation (\ref{strong}) in the previous section. We have found that the imaginary part of optical potential has a important role for the repulsive shift in kaonic atoms. Hence next we examine the optical potential which has a non-linear term, and and confirm whether it provides the repulsive shift not by the level repulsion of nuclear states but by the imaginary part of the potential. 

The phenomenological potential reported in Ref.~\cite{Friedman:1994hx} has non-linear density terms and is given by $V_{\mathrm{opt}}(r) = -4\pi\eta a_{\mathrm{eff}}(\rho_N)\rho_N/2\mu$ and $a_{\mathrm{eff}}(\rho_N) = (-0.15+0.62i) + (1.66-0.04i)(\rho_N/\rho_0)^{0.21}\ \mathrm{fm}$
where $a_{\mathrm{eff}}(\rho_N)$ is a density dependent effective scattering length and $\eta = 1 + m_K/m_N$. This potential amounts to $( -190-i 80)~\mathrm{MeV}$ at the nuclear center.
The Real part of the potential is almost proportional to the non-linear term and the imaginary part is almost proportional to the linear term. 

Here we take the following form to study the non-linear term effect :
\begin{align}
  V_{\mathrm{opt}}(r) = -V_0\qty(\frac{\rho_N(r)}{\rho_0})^\alpha - iW_0\qty(\frac{\rho_N(r)}{\rho_0}),
\end{align}
where $\alpha$ is a parameter which characterizes the non-linear effect. We assume the real part of this potential is proportional to the non-linear term and the imaginary part is to the linear term.
We determine the parameters $V_0$ and $W_0$ again from the observed shift and width for the non-linear parameter $\alpha$ changed from $\alpha=1$ (Potential~1) to $\alpha=1.231$.
We carry out the calculation to find the parameters $V_0$ and $W_0$ for the Cu kaonic atom for $\ell=3$ and summarize the result in Table \ref{tab:NL}.
This shows that the potential with $\alpha = 1.231$ has very similar potential parameters with the phenomenological potential obtained in Ref.~\cite{Friedman:1994hx}. It is also found that potentials with stronger nonlinearity have a deeper real part and a smaller imaginary part. It is interesting that this potential is continuously connected to Potential~1 with the parameter $\alpha$. This fact implies that the phenomenological potential in Ref.~\cite{Friedman:1994hx} may have similar properties to Potential~1. 

\begin{table}
  \caption{Determined potential parameters for each $\alpha$. The Cu$^1$ data are used to determine the parameter. The $\alpha$ parameter characterizes the nonlinearity of the density dependence. }
  \label{tab:NL}
  \begin{ruledtabular}
    \begin{tabular*}{8.6cm}{@{\extracolsep{\fill}}ccc}
      $\alpha$     & $V_0$[MeV]    & $W_0$[MeV]  \\ \hline
      1.000           & 79.5        & 114.5     \\
      1.033           & 90.0        & 109.0    \\
      1.066           & 101.0       & 105.0    \\
      1.099           & 114.0       & 99.0     \\
      1.132           & 127.0       & 94.0    \\
      1.165           & 141.0       & 89.0    \\
      1.198           & 157.0       & 81.0    \\
      1.231           & 174.0       & 70.0     \\
    \end{tabular*}
  \end{ruledtabular}
\end{table}

In order to see the nature of the obtained potential, we calculate the energy spectrum of the kaonic nucleus with $\ell=3$ using the following potential:
\begin{align}
  \label{pot_NL1}
  V_{\mathrm{opt}}(r) & = -174.0\qty(\frac{\rho_N(r)}{\rho_0})^{1.231} - i\ 70.0\qty(\frac{\rho_N(r)}{\rho_0})\ \mathrm{MeV}.
\end{align}

This is the most similar to the phenomenological potential given in Ref.~\cite{Friedman:1994hx}. This potential provides nuclear states for $\ell = 3$ with $-E_B-i \Gamma/2 = (-52.1-i 85.1/2)$ MeV as the ground state and with $-E_B-i \Gamma/2 = (+12.8-i 69.5/2)$ MeV as a radial excited state with a negative binding energy and a large width. The ground state has such a deep binding energy that the nuclear state hardly influences the atomic states and its absorption width is also large. The excited state is a resonance with a negative binding energy trapped by the centrifugal barrier and is sitting above the atomic states. This feature is different from Potentials 2 and 3. The potential (\ref{pot_NL1}) has very similar characteristics to Potential~1, which provides the repulsive shift by the effect of the imaginary part. 

We discuss the universality of the potential (\ref{pot_NL1}). We calculate the energy shifts and absorption widths for the various kaonic atoms using the optical potentials (\ref{pot_NL1}). We confirm that the potential reproduces the observed data seen in Table \ref{tab:Katomdata} within the experimental error as found for Potential~1 ($V_{\mathrm{opt}}(r) = -(79.5 + 114.5i)\qty(\rho_N(r)/\rho_0)\ \mathrm{MeV}$) in the previous section. This is a natural result because the potential (\ref{pot_NL1}) is similar to the phenomenological potential is Ref~\cite{Friedman:1994hx} and it has already universality to reproduce a wide range of the observed kaonic atoms.

\begin{figure}[t]
  \includegraphics[width=0.45\textwidth]{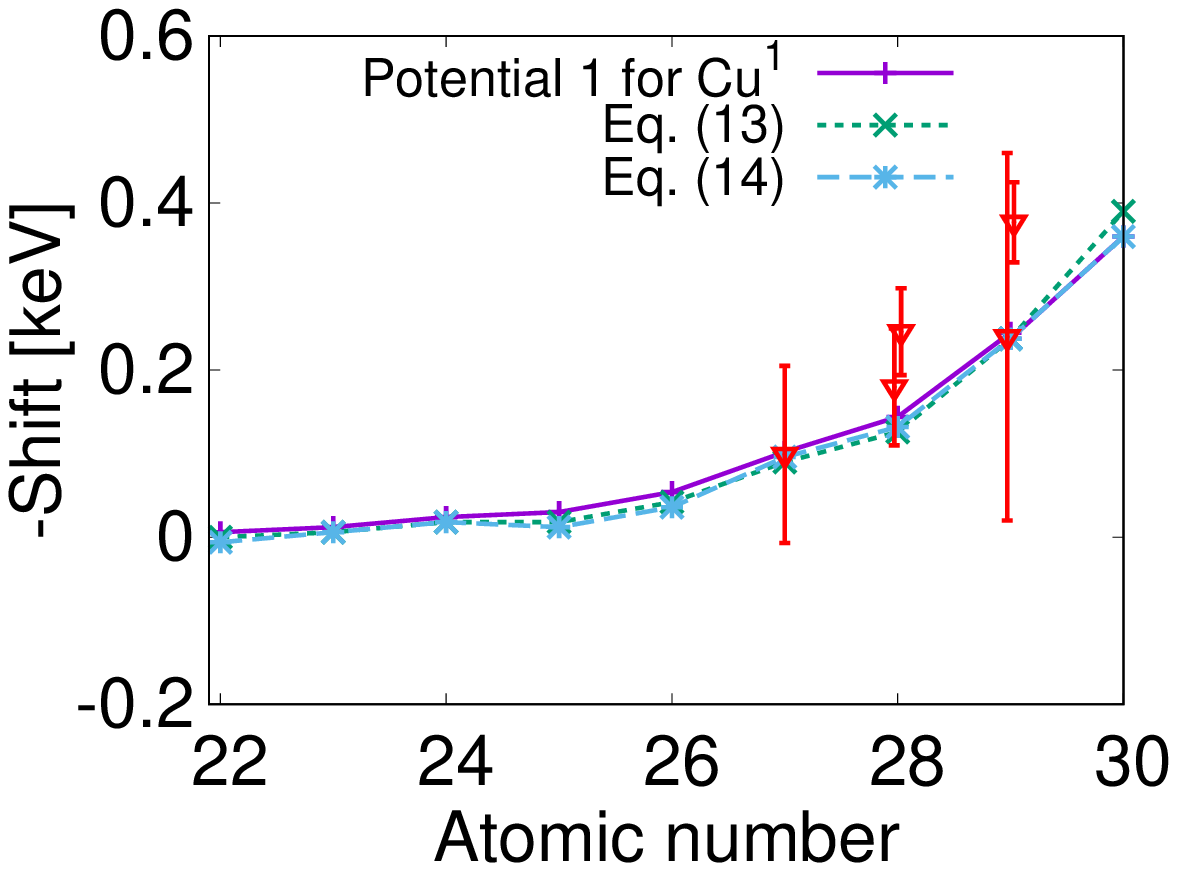}
  \includegraphics[width=0.45\textwidth]{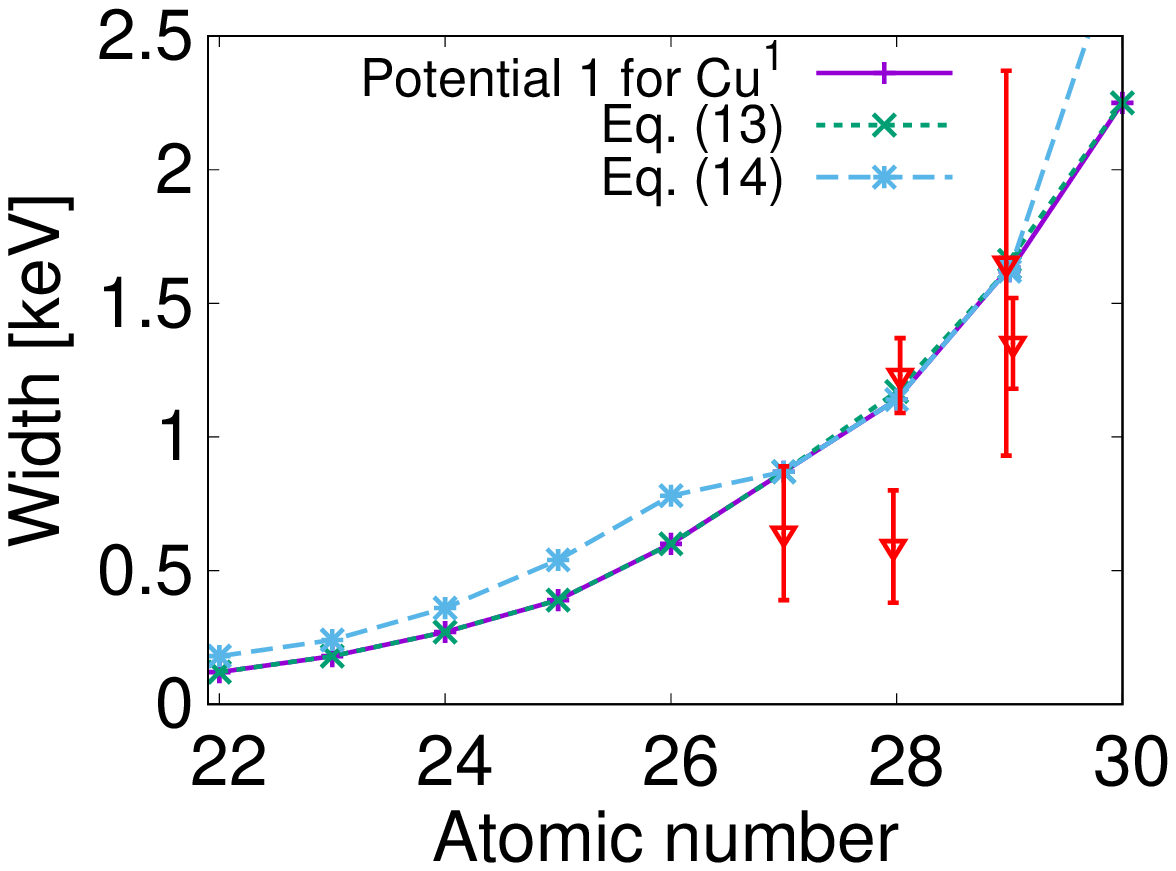}
  \caption{Energy shifts (upper plot) and absorption width (lower plot) of the kaonic atoms with $\ell = 3$ calculated with Potential~1 for Cu$^1$, Eq.~(\ref{pot_NL1}) and Eq.~(\ref{pot_unfolded}). The solid, dashed and dotted lines denote the results obtained by Potential~1, Eq.~(\ref{pot_NL1}) and Eq.~(\ref{pot_unfolded}), respectively. The points with error bars are the experimental data. }
  \label{fig:4f_NL}
\end{figure}

In Fig.~\ref{fig:4f_NL} we plot the energy shifts and the absorption widths of the kaonic atoms with the last orbit 4f calculated by the potential (\ref{pot_NL1}) and Potential~1 for Cu$^1$. This figure shows that the potential (\ref{pot_NL1}) provides almost same atomic spectrum with Potential~1 and is consistent with the observation. Therefore, even though potentials with nonlinearity have a deeper real part, nuclear states play no significant role for the repulsive energy shift of the atomic state and have similar features with Potential~1, in which the imaginary part of the optical potential works essentially for the repulsive shift.

\subsection{Another nuclear density parameter}
In this section, we study the optical potential with another nuclear density parameter, which is originally used in Ref.~\cite{Friedman:1994hx} and referred to as ``Unfolded'' there. Here we list the density parameters for the Co, Ni and Cu nuclei in Table~\ref{tab:Unfolded}. These density parameters have a bit larger radius $R_B$ and a smaller diffuseness $a$. Using the ``Unfolded'' nuclear density parameter, we calculate the optical potential parameters for Cu to reproduce the Cu$^1$ datum in the linear density approximation and obtain the optical potential
\begin{align}
  \label{pot_unfolded}
  V_{\mathrm{opt}}(r) & = -(156.0+ i\ 122.0) \qty(\frac{\rho_N(r)}{\rho_0})\ \mathrm{MeV}.
\end{align}
This potential again have a larger real part, which has a almost double depth in comparison with Potential~1, while the imaginary part is as large as Potential~1. In Fig.~\ref{fig:4f_NL}, we compare the energy shifts and absorption widths calculated by the potential~(\ref{pot_unfolded}) and Potential~1 for Cu$^1$. Again we find that the potential~(\ref{pot_unfolded}) and Potential~1 give the almost same atomic spectrum and the potential~(\ref{pot_unfolded}) is consistent with the experiments for other nuclei than Cu, even though the real parts of these optical potentials are quite different. This implies that the potentials that successfully reproduce the observed energy shifts and absorption width of kaonic atoms widely have a large imaginary part and it is responsible for the repulsive shift, no matter how the real part of the optical potential is large. 

\begin{table}
  \caption{Unfolded nuclear parameters taken from Ref.~\cite{Friedman:1994hx}}
  \label{tab:Unfolded}
  \begin{ruledtabular}
    \begin{tabular*}{8.6cm}{@{\extracolsep{\fill}}ccc}
      Nuclei     & $R_B$[fm]   & $a$[fm]  \\ \hline
      Co         & 4.124       & 0.504    \\
      Ni         & 4.134       & 0.500    \\
      Cu         & 4.243       & 0.501    \\
    \end{tabular*}
  \end{ruledtabular}
\end{table}

\section{Conclusion}
\label{sec:6}
In this paper, we have clarified the origin of the repulsive energy shifts for the last orbits of the kaonic atoms from pure Coulomb spectrum. The repulsive shifts are observed universally in all of the observed kaonic atoms. Having assumed the optical potentials to be propositional to nuclear density $\rho_N$, we have introduced two parameters for the real and imaginary parts of the optical potential. We have determined these potential parameters so as to reproduce the observed repulsive shift and absorption width for each kaonic atom datum. We have obtained two kinds of potentials for each datum; in the first case (Potential~1), the potential has a large imaginary part and the imaginary part works repulsively, and in the second case (Potentials~2 and 3), the potential has a substantially large real part that produces nuclear bound states with the same angular momentum to the atomic state and the nuclear states push up repulsively the atomic state. Having confirmed the universality of these potentials, we have found that Potential~1, which has a large imaginary part, explains the experimental data globally, but Potentials~2 and 3 can explain only limited kaonic atoms. Therefore, we conclude that the imaginary part of the optical potential plays an important role in the repulsive shift of kaonic atoms. This result implies that it would be somewhat difficult to observe nuclear states in heavier nuclei because of large nuclear absorption.

\begin{acknowledgments}
  The work of D.J. was partly supported by Grants-in-Aid for Scientific Research from Japan Society for the Promotion of Science (JSPS) (17K05449).
  The work of N.I. was partly supported by JSPS Overseas Research Fellowships and JSPS KAKENHI Grant Number JP19K14709.
  The work of J. Y. was partly supported by JSPS KAKENHI Grant Number JP18K13545.
  The work of S.H. was partly supported by Grants-in-Aid for Scientific Research from JSPS (16K05355). 
\end{acknowledgments}

\end{document}